\documentclass[a4paper,12pt]{article}
\usepackage[T1]{fontenc}
\usepackage[margin=1.075in]{geometry}
\usepackage{amsmath,amsthm,amssymb,mathtools}
\usepackage{bm}
\usepackage{array}
\usepackage{float}
\usepackage[colorlinks,citecolor=blue,urlcolor=blue,linkcolor=blue,filecolor=blue,backref=page]{hyperref}
\usepackage[nottoc,numbib]{tocbibind}
\usepackage[toc,page,title]{appendix}
\usepackage{authblk}
\usepackage{enumitem}
\usepackage{graphicx}
\usepackage[figuresright]{rotating}
\usepackage{epstopdf}
\usepackage{nicefrac}
\usepackage{subcaption}
\usepackage{multirow}
\usepackage[linesnumbered,ruled]{algorithm2e}
\usepackage{setspace}
\usepackage[authoryear,semicolon,round]{natbib}
\usepackage{listings}
\def\given{\,|\,}
\newcommand{\argmin}{{\operatorname{arg}\operatorname{min}}}
\allowdisplaybreaks
\usepackage{etoolbox}\AtBeginEnvironment{algorithmic}{\small}
\providecommand{\keywords}[1]{\footnotesize{\textbf{Keywords:} #1}}
\makeatletter
\newcommand{\mathleft}{\@fleqntrue\@mathmargin1.5em}
\newcommand{\mathcenter}{\@fleqnfalse}
\makeatother
\makeatletter
\lst@InstallKeywords k{attributes}{attributestyle}\slshape{attributestyle}{}ld
\makeatother
\newcolumntype{K}[1]{>{\centering\arraybackslash}p{#1}}
\newcommand{\minsub}[1]{\underset{#1}{\operatorname{min}}\;}

\usepackage[framemethod=TikZ]{mdframed}
\usepackage{tikz}
\usepackage{ctable}
\usetikzlibrary{bayesnet}
\usetikzlibrary{shadows}
\usepackage{environ}
\usepackage{varwidth}
\usepackage{ctable}

\newlength{\MyMdframedWidthTweak}%
\NewEnviron{reminder}[1][]{%
	\setlength{\MyMdframedWidthTweak}{\dimexpr%
		+\mdflength{innerleftmargin}
		+\mdflength{innerrightmargin}
		+\mdflength{leftmargin}
		+\mdflength{rightmargin}
	}%
	\savebox0{%
		\begin{varwidth}{\dimexpr0.9\linewidth-\MyMdframedWidthTweak\relax}%
			\BODY
		\end{varwidth}%
	}%
	\begin{mdframed}[
		backgroundcolor=gray!20, 
		shadow=true, 
		align=center,
		shadowsize=4pt,
		roundcorner=5pt,
		userdefinedwidth=\dimexpr\wd0+\MyMdframedWidthTweak\relax, 
		#1]
		\usebox0
	\end{mdframed}
}

\title{\textbf{Gaussian Parsimonious Clustering Models}\\ \textbf{with Covariates and a Noise Component}\vspace{-0.5em}}
\author{\vspace{-0.75ex}Keefe Murphy\textsuperscript{1} \qquad Thomas Brendan Murphy\textsuperscript{2,3} \\\vspace{-0.5ex} \href{mailto:me@keefe.murphy@mu.ie}{\small{keefe.murphy@mu.ie}}~ \href{mailto:me@brendan.murphy@ucd.ie}{\small{brendan.murphy@ucd.ie}}\vspace{-0.5em}}

\affil[]{{\footnotesize\textsuperscript{1}}~\footnotesize{Department of Mathematics and Statistics, Maynooth University}\\\vspace{-0.33ex}{\footnotesize\textsuperscript{2}}~\footnotesize{School of Mathematics and Statistics, University College Dublin}\\\vspace{-0.33ex} {\footnotesize\textsuperscript{3}}~\footnotesize{Insight Centre for Data Analytics, University College Dublin}}

\date{}

\begin{document}
	
\vspace*{-4.5em}
\begin{center}
\begin{reminder}
	{\centering \color{red} \textbf{\textsf{\small This is a preprint. The revised version of this paper is published as}}\\}
	\vspace{3pt}
	{\small \noindent K. Murphy and T.\,B. Murphy (2020) %
		{``Gaussian parsimonious clustering models\linebreak with covariates and a noise component''.} %
		\textit{Advances in Data Analysis and\linebreak Classification}, 14(2): 293--325.\hfill [\href{https://doi.org/10.1007/s11634-019-00373-8}{\sf doi: 10.1007/s11634-019-00373-8}].}
\end{reminder}
\end{center}
\vspace{-4em}
{\let\newpage\relax\maketitle}
\vspace{-3.75em}

\begin{abstract}
	We consider model-based clustering methods for continuous, correlated data that account for external information available in the presence of mixed-type fixed covariates by proposing the MoEClust suite of models. These models allow different subsets of covariates to influence the component weights and/or component densities by modelling the parameters of the mixture as functions of the covariates. A familiar range of constrained eigen-decomposition parameterisations of the component covariance matrices are also accommodated. This paper thus addresses the equivalent aims of including covariates in Gaussian parsimonious clustering models and incorporating parsimonious covariance structures into all special cases of the Gaussian mixture of experts framework. The MoEClust models demonstrate significant improvement from both perspectives in applications to both univariate and multivariate data sets. Novel extensions to include a uniform noise component for capturing outliers and to address initialisation of the EM algorithm, model selection, and the visualisation of results are also proposed.\bigskip
	
	\noindent\keywords{covariates, EM algorithm,  mixtures of experts, model-based clustering, multivariate response, noise component,  parsimony.}
\end{abstract}
\vspace{-1.25em}

\section{Introduction}
\label{sec:Intro}
In many analyses using the standard mixture model framework, a clustering method~is typically implemented on the outcome variables only. Reference is not made to the associated covariates until the structure of the produced clustering is investigated in light of the information present in the covariates. Therefore, interpretations of~the values of the model parameters within each component are guided by covariates that are~not actually used in the construction of the clusters. It is desirable to have covariates incorporated into the clustering process and not only into the interpretation~of the clustering structure and model parameters, thereby making them endogenous rather~than exogenous to the clustering model. This both informs the construction of the clusters and provides richer insight into the type of observation which characterises each cluster.

When each observation consists of a response variable $\mathbf{y}_i$ on which the~\mbox{clustering~is} based and covariates $\mathbf{x}_i$ there are, broadly speaking, two main approaches in the literature to having covariates guide construction of the clusters, neatly summarised by\linebreak \citet{Lamont2016} and compared in \citet{Ingrassia2012}. Letting $\mathbf{z}_i$ denote the latent cluster membership indicator vector, where $z_{ig}=1$ if observation $i$ belongs to cluster $g$ and $z_{ig}=0$ otherwise, the first approach assumes that $\mathbf{z}_i$ affects the distribution of $\mathbf{x}_i$. In probabilistic terms,
this means to replace the actual group-specific conditional distribution $f\big(\mathbf{y}_i \given \mathbf{x}_i, z_{ig} = 1\big)\Pr\negthinspace\big(z_{ig} = 1\big)$ with $f\big(\mathbf{y}_i | \mathbf{x}_i, z_{ig} = 1\big)f\big(\mathbf{x}_i \given z_{ig} = 1\big)\Pr\negthinspace\big(z_{ig} = 1\big)$. The name `cluster-weighted model' (CWM) is frequently given to this approach, e.g. \citet{Dang2017} and \citet{Ingrassia2015}; the latter provides a recent extension allowing for mixed-type covariates, with a further generalisation presented in \citet{Punzo2016}. Noting the use of the alternative term `mixtures of regressions with \emph{random} covariates' to describe CWMs (e.g. \citealt{Hennig2000}) provides opportunity to clarify that the remainder of this paper focuses on the second approach, with \emph{fixed} potentially mixed-type covariates affecting  cluster membership via $f\big(\mathbf{y}_i \given \mathbf{x}_i, z_{ig} = 1\big)\Pr\negthinspace\big(z_{ig} = 1 \given \mathbf{x}_i\big)$.

This is achieved using the mixture of experts (MoE) paradigm \citep{Dayton1988, Jacobs1991} in which the parameters of the mixture are modelled as~functions of fixed, potentially mixed-type covariates. We present, for finite mixtures of~multivariate, continuous, correlated responses, a unifying framework combining all~of the special cases of the Gaussian MoE model with the flexibility afforded by the covariance constraints in the Gaussian parsimonious clustering model (GPCM) family\linebreak \citep{Banfield1993, Celeux1995}. This has, to date, been lacking for all but mixtures of regressions and mixtures of regressions with concomitant~\mbox{variables} where the same covariates enter both parts of the model \citep{Dang2015}.

Parsimony is obtained in GPCMs by imposing constraints on the elements of an eigen-decomposition of the component covariance matrices. For MoE models, reducing the number of covariance parameters in this manner can help offset the number~of regression parameters introduced by covariates, which is particularly advantageous when model selection is conducted using information criteria with penalty terms involving parameter counts. The main contribution of this paper is the development of a framework combining GPCM constraints with all of the special cases of the Gaussian MoE framework whereby different subsets of covariates can enter either, neither, or both the component densities and component weights. We also consider the special cases of the MoE framework for univariate response data with equal and unequal variance across components. Thus, this paper addresses the aim of incorporating potentially mixed-type covariates into the GPCM family and the equivalent aim of bringing GPCM covariance constraints into the Gaussian MoE framework, by proposing the MoEClust model family. The name MoEClust comes from the interest in employing MoE models chiefly for clustering purposes. From both perspectives, MoEClust models show significant improvement in applications to both univariate and multivariate response~data. 

Other novel contributions include the addition of a noise component for capturing outlying observations, proposed solutions to initialising the EM algorithm sensibly, addressing issues of model selection, and a means for visualising the results of MoEClust models. We also expand the number of special cases in the MoE framework from~four to six, by considering more parsimonious counterparts to the standard mixture model and the mixture of regressions by constraining the mixing proportions. In addition, a software implementation of the full suite of MoEClust models is provided by the associated \textsf{R} package \texttt{MoEClust} \citep{MoEClustR2021}, with which all results were obtained, which is available from \href{https://www.r-project.org}{\texttt{https://www.r-project.org}} \citep{R2021}. The syntax of the popular \texttt{mclust} package \citep{Scrucca2016} is closely mimicked, with formula interfaces for specifying covariates in the gating and/or expert networks.

The structure of the paper is as follows. For both Gaussian mixtures of experts and MoEClust models, the modelling frameworks and inferential procedures are described, respectively, in Section \ref{sec:Modelling} and Section \ref{sec:Inference}. Section \ref{sec:noise} describes the addition of a noise component for capturing outliers. Section \ref{sec:Performance} discusses proposals for addressing some practical issues affecting performance, namely the initialisation of the EM algorithm used to fit the models (Section \ref{sec:init}), and issues around model selection (Section \ref{sec:selection}). The performance of the proposed models is illustrated in Section \ref{sec:Results} with applications to univariate response CO\textsubscript{2} emissions data (Section \ref{sec:CO2}) and multivariate response data from the Australian Institute of Sports (Section \ref{sec:AIS}). Finally, the paper concludes with a brief discussion in Section \ref{sec:Discussion}, with some additional results deferred to the Appendices.

\section[Modelling]{Modelling}
\label{sec:Modelling}
This section builds up the MoEClust models by first describing the mixture of experts (MoE) modelling framework in Section \ref{sec:MoE} --- elaborating on the special cases of the MoE model in Section \ref{sec:family} --- and then extending to the family of MoEClust models comprising Gaussian mixture of experts models with  parsimonious covariance structures from the GPCM family in Sections \ref{sec:parsimonious} and \ref{sec:family2}. Finally, a brief review of existing models and software is given in Section \ref{sec:software}.

\subsection[Mixtures of Experts]{Mixtures of Experts}
\label{sec:MoE}
The mixture of experts model \citep{Dayton1988,Jacobs1991} extends the mixture model used to cluster response data $\mathbf{y}_i$ by allowing the parameters~of the model for observation $i$ to depend on covariates $\mathbf{x}_i$. An independent sample of response/outcome variables of dimension $p$, denoted by $\mathbf{Y}=\left(\mathbf{y}_1,\ldots,\mathbf{y}_n\right)$, is modelled by a $G$-component finite mixture model where the model parameters depend on the associated covariate inputs $\mathbf{X}=\left(\mathbf{x}_1,\ldots,\mathbf{x}_n\right)$ of dimension $d$. The MoE model is often referred to as a conditional mixture model \citep{Bishop2006} because, given the set of covariates $\mathbf{x}_i$, the distribution of the response variable $\mathbf{y}_i$ is a finite mixture model:
\[f\big(\mathbf{y}_i \given \mathbf{x}_i\big) = \sum_{g=1}^{G}\tau_g\negthinspace\left(\mathbf{x}_i\right)f\big(\mathbf{y}_i\given\boldsymbol{\theta}_g\left(\mathbf{x}_i\right)\big).\]
\noindent Each component is modelled by a probability density function $f\big(\mathbf{y}_i\given\boldsymbol{\theta}_g\left(\mathbf{x}_i\right)\big)$ with component-specific parameters $\boldsymbol{\theta}_g\left(\mathbf{x}_i\right)$ and mixing proportions $\tau_g\negthinspace\left(\mathbf{x}_i\right)$ which are only allowed to depend on covariates when $G\geq 2$.~As usual, $\tau_g\negthinspace\left(\mathbf{x}_i\right) > 0$ and $\sum_{g=1}^{G}\tau_g\negthinspace\left(\mathbf{x}_i\right)=\nolinebreak1$.

The MoE framework facilitates flexible modelling. While the response variable~$\mathbf{y}_i$~is modelled via a finite mixture, model parameters are modelled as functions of related covariates $\mathbf{x}_i$ from the context under study. Both the mixing proportions and the parameters of component densities can depend on $\mathbf{x}_i$. The terminology used to describe MoE models in the machine learning literature often refers to the component densities $f\big(\mathbf{y}_i\given\boldsymbol{\theta}_g\left(\mathbf{x}_i\right)\big)$ as `experts' or~the `expert network', and to the mixing proportions $\tau_g\negthinspace\left(\mathbf{x}_i\right)$ as `gates' or the `gating network', hence the nomenclature \emph{mixture of experts}. Given that covariates can be continuous and/or categorical with multiple levels, we let $d + 1$ denote the number of columns in the corresponding design matrices, accounting also for the intercept term, in contrast to the number of covariates $r$, with $d \geq r$.

In the original formulation of the MoE model for continuous data \citep{Jacobs1991}, the mixing proportions (gating network) are modelled using multinomial logistic regression (MLR), though this need not strictly be the case; \citet{Geweke2007} impose a multinomial probit structure here instead, while \citet{Xu1994} effectively obtain a CWM (albeit with strictly continuous covariates) by assuming Gaussian gating functions. The mixture components~(expert networks) are generalised linear models (GLM; \citealp{McCullagh1983}). Thus,\vspace{-1ex}
\begin{align}
\widehat{\tau}_g\negthinspace\left(\mathbf{x}_i\right) &= \frac{\exp\negthinspace\big( \widetilde{\mathbf{x}}_i\widehat{\boldsymbol{\beta}}_g\big)}{\sum_{h=1}^{G}\exp\negthinspace\big(\widetilde{\mathbf{x}}_i\widehat{\boldsymbol{\beta}}_h\big)},\label{eq:tautheta}
\shortintertext{and}
\widehat{\boldsymbol{\theta}}_g\negthinspace\left(\mathbf{x}_i\right) &= \left\{\psi\big(\widetilde{\mathbf{x}}_i\widehat{\boldsymbol{\gamma}}_g\big), \widehat{\boldsymbol{\Sigma}}_g\right\},\label{eq:explink}\vspace{-1ex}
\end{align}
\noindent for some link function $\smash{\psi\negthinspace\left(\cdot\right)}$, with a collection of parameters in the component densities (comprising a $\smash{\left(d + 1\right)\times p}$ matrix of expert network regression parameters $\smash{\widehat{\boldsymbol{\gamma}}_g}$ and the $p\times p$ component covariance matrix $\smash{\widehat{\boldsymbol{\Sigma}}_g}$), a $\smash{\left(d + 1\right)}$-dimensional vector of regression parameters $\smash{\widehat{\boldsymbol{\beta}}}_g$ in the gates in \eqref{eq:tautheta}, and $\smash{\widetilde{\mathbf{x}}_i = \left(1, \mathbf{x}_i\right)}$. Note that expert network covariates influence only the component means, and not the component covariance matrices. Henceforth, we restrict our attention to continuous outcome variables as per the GPCM family. Therefore, component densities are assumed to be the $p$-variate Gaussian $\smash{\phi\negthinspace\left(\mathbf{y}_i\given\cdot\right)}$, and the link function $\smash{\psi\negthinspace\left(\cdot\right)}$ in \eqref{eq:explink} is simply the identity, such that covariates are linearly related to the response variables, i.e.\vspace{-1ex}
\begin{align}
f\big(\mathbf{y}_i \given \mathbf{x}_i\big) = \sum_{g=1}^{G}\tau_g\negthinspace\left(\mathbf{x}_i\right)\phi\big(\mathbf{y}_i\given\boldsymbol{\theta}_g\left(\mathbf{x}_i\right) = \big\{\widetilde{\mathbf{x}}_i\boldsymbol{\gamma}_g, \boldsymbol{\Sigma}_g\big\}\big).\label{eq:fullMoE}\vspace{-1ex}
\end{align}

\subsubsection[The MoE Family of Models]{The MoE Family of Models}
\label{sec:family}
It is possible that some, none, or all model parameters depend on the covariates. This leads to the four special cases of the Gaussian MoE framework shown in Figure \ref{Plot:GraphicalRep}, with the following interpretations, due to \citet{Gormley2011}:\enlargethispage{\baselineskip}
\begin{enumerate}[label=(\alph*)]
	\itemsep 0.25em
	\item in the \emph{mixture model} the distribution of $\smash{\mathbf{y}_i}$ depends on the latent cluster membership variable $\smash{\mathbf{z}_i}$, the distribution of $\smash{\mathbf{z}_i}$ is independent of the covariates $\smash{\mathbf{x}_i}$, and $\smash{\mathbf{y}_i}$ is independent of $\smash{\mathbf{x}_i}$ conditional on $\smash{\mathbf{z}_i\colon f\big(\mathbf{y}_i\big) = \sum_{g=1}^{G}\tau_g \phi\big(\mathbf{y}_i\given\boldsymbol{\theta}_g=\big\{\boldsymbol{\mu}_g,\boldsymbol{\Sigma}_g\big\}\big).}$\label{eq:MoEmix}
	\item in the \emph{expert network MoE model} the distribution of $\smash{\mathbf{y}_i}$ depends on the covariates $\smash{\mathbf{x}_i}$ and the latent cluster membership variable $\smash{\mathbf{z}_i}$, and the distribution of $\smash{\mathbf{z}_i}$ is independent of $\smash{\mathbf{x}_i\colon f\big(\mathbf{y}_i\given \mathbf{x}_i\big) = \sum_{g=1}^{G}\tau_g \phi\big(\mathbf{y}_i\given\boldsymbol{\theta}_g\big(\mathbf{x}_i\big) = \big\{\widetilde{\mathbf{x}}_i\boldsymbol{\gamma}_g, \boldsymbol{\Sigma}_g\big\}\big).}$\label{eq:MoEexp}
	\item in the \emph{gating network MoE model} the distribution of $\smash{\mathbf{y}_i}$ depends on the latent cluster membership variable $\smash{\mathbf{z}_i}$, $\smash{\mathbf{z}_i}$ depends on the covariates $\smash{\mathbf{x}_i}$, and $\smash{\mathbf{y}_i}$ is independent of $\smash{\mathbf{x}_i}$ conditional on $\smash{\mathbf{z}_i\colon f\big(\mathbf{y}_i\given \mathbf{x}_i\big) = \sum_{g=1}^{G}\tau_g\big(\mathbf{x}_i\big) \phi\big(\mathbf{y}_i\given\boldsymbol{\theta}_g=\big\{\boldsymbol{\mu}_g,\boldsymbol{\Sigma}_g\big\}\big).}$\label{eq:MoEgate}
	\item in the \emph{full MoE model}, given by \eqref{eq:fullMoE}, the distribution of $\smash{\mathbf{y}_i}$ depends on both the covariates $\smash{\mathbf{x}_i}$ and on the latent cluster membership variable $\smash{\mathbf{z}_i}$, and the distribution of the latent variable $\smash{\mathbf{z}_i}$ depends in turn on the covariates $\smash{\mathbf{x}_i}$. \label{eq:MoEfull}
\end{enumerate}
For models \ref{eq:MoEgate} and \ref{eq:MoEfull}, $\smash{\mathbf{z}_i}$ has a multinomial distribution with a single trial and~probabilities equal to $\smash{\tau_g\negthinspace\left(\mathbf{x}_i\right)}$. The full MoE model thus has the following latent variable representation: $\smash{\big(\mathbf{y}_i\given \mathbf{x}_i,z_{ig} = 1\big)\sim \phi\big(\mathbf{y}_i\given\boldsymbol{\theta}_g\left(\mathbf{x}_i\right) = \big\{\widetilde{\mathbf{x}}_i\boldsymbol{\gamma}_g, \boldsymbol{\Sigma}_g\big\}\big),\Pr\negthinspace\left(z_{ig} = 1\given \mathbf{x}_i\right) = \tau_g\negthinspace\left(\mathbf{x}_i\right)}$.
\begin{figure}[H]
	\centering
	\begin{subfigure}[tc]{.49\linewidth}
		\centering
		\tikz[baseline,remember picture]{
			\node[obs, fill=green!30] (X) {$\mathbf{X}$};%
			\node[latent,left=of X,xshift=0cm] (Z) {$\mathbf{Z}$}; %
			\node[obs,below=of Z, fill=green!30] (Y) {$\mathbf{Y}$}; %
			\node[latent,left=of Z,xshift=0cm, draw=none] (tau) {$\boldsymbol{\tau}$}; %
			\node[latent,left=of Y,xshift=0cm, draw=none] (theta) {$\boldsymbol{\theta} = \left\{\boldsymbol{\mu}, \boldsymbol{\Sigma}\right\}$}; %
			\plate [yshift=.05cm, color=red, thick, minimum width=10ex] {plate1} {(X)(Y)(Z)} {}; %
			\edge {tau} {Z};
			\edge {theta} {Y};
			\edge {Z} {Y}  }
		\captionsetup{justification=centering}
		\caption{Mixture model.\label{Subplot:MixtureModel}}
	\end{subfigure}
	\begin{subfigure}[tc]{.49\linewidth}
		\centering
		\tikz[baseline,remember picture]{
			\node[obs, fill=green!30] (X) {$\mathbf{X}$};%
			\node[latent,left=of X,xshift=0cm] (Z) {$\mathbf{Z}$}; %
			\node[obs,below=of Z, fill=green!30] (Y) {$\mathbf{Y}$}; %
			\node[latent,left=of Z,xshift=0cm, draw=none] (tau) {$\boldsymbol{\tau}$}; %
			\node[latent,left=of Y,xshift=0cm, draw=none] (theta) {$\boldsymbol{\theta} = \left\{\boldsymbol{\gamma}, \boldsymbol{\Sigma}\right\}$}; %
			\plate [yshift=.05cm, color=red, thick, minimum width=10ex] {plate1} {(X)(Y)(Z)} {}; %
			\edge {tau} {Z};
			\edge {theta} {Y};
			\edge {Z} {Y};
			\edge {X} {Y}  }
		\captionsetup{justification=centering}
		\caption{Expert network MoE model.\label{Subplot:ExpertModel}}
	\end{subfigure}
	\begin{subfigure}[tc]{.49\linewidth}
		\centering
		\vspace{\baselineskip}
		\tikz[baseline,remember picture]{
			\node[obs, fill=green!30] (X) {$\mathbf{X}$};%
			\node[latent,left=of X,xshift=0cm] (Z) {$\mathbf{Z}$}; %
			\node[obs,below=of Z, fill=green!30] (Y) {$\mathbf{Y}$}; %
			\node[latent,left=of Z,xshift=0cm, draw=none] (beta) {$\boldsymbol{\beta}$}; %
			\node[latent,left=of Y,xshift=0cm, draw=none] (theta) {$\boldsymbol{\theta} = \left\{\boldsymbol{\mu}, \boldsymbol{\Sigma}\right\}$}; %
			\plate [yshift=.05cm, color=red, thick, minimum width=10ex] {plate1} {(X)(Y)(Z)} {}; %
			\edge {beta} {Z};
			\edge {theta} {Y};
			\edge {Z} {Y};
			\edge {X} {Z}  }
		\captionsetup{justification=centering}
		\caption{Gating network MoE model.\label{Subplot:GatingModel}}
	\end{subfigure}
	\begin{subfigure}[tc]{.49\linewidth}
		\centering
		\vspace{\baselineskip}
		\tikz[baseline,remember picture]{
			\node[obs, fill=green!30] (X) {$\mathbf{X}$};%
			\node[latent,left=of X,xshift=0cm] (Z) {$\mathbf{Z}$}; %
			\node[obs,below=of Z, fill=green!30] (Y) {$\mathbf{Y}$}; %
			\node[latent,left=of Z,xshift=0cm, draw=none] (beta) {$\boldsymbol{\beta}$}; %
			\node[latent,left=of Y,xshift=0cm, draw=none] (theta) {$\boldsymbol{\theta} = \left\{\boldsymbol{\gamma}, \boldsymbol{\Sigma}\right\}$}; %
			\plate [yshift=.05cm, color=red, thick, minimum width=10ex] {plate1} {(X)(Y)(Z)} {}; %
			\edge {beta} {Z};
			\edge {theta} {Y};
			\edge {Z} {Y};
			\edge {X} {Z};
			\edge {X} {Y};  }
		\captionsetup{justification=centering}
		\caption{Full MoE model.\label{Subplot:FullModel}}
	\end{subfigure}
	\caption{The graphical model representation of the mixture of experts models. The differences between the special cases are due to the presence or absence of edges between the covariates $\mathbf{X}$ and the latent variables $\mathbf{Z}$ and/or response variables $\mathbf{Y}$. Note that different subsets of the covariates in $\mathbf{X}$ can enter these two different parts of the full MoE model in (\subref{Subplot:FullModel}).}\label{Plot:GraphicalRep}
\end{figure}
\vspace{-0.5em}
The MoE family can be expanded further, from four to six special cases, by considering the models in \ref{eq:MoEmix} and \ref{eq:MoEexp}, under which covariates do not enter the gating network, by constraining the mixing proportions to be equal across components, i.e. $\tau_g=\nicefrac{1}{G}\:\:\forall\:\:g$. This leads, respectively, to the \emph{equal mixing proportion mixture model} and \emph{equal mixing proportion expert network MoE model}. Such models are more parsimonious than their counterparts with unconstrained $\boldsymbol{\tau}$, as they require estimation of $G-1$ fewer parameters. Note that the size of a cluster is proportional to $\tau_g$, which is distinct from its volume \citep{Celeux1995}. Thus, situations where $\tau_{ig}=\tau_g\negthinspace\left(\mathbf{x}_i\right)$, $\tau_{ig}=\tau_g$, or $\tau_{ig}=\nicefrac{1}{G}$ can all be accommodated. The six special cases of this MoE framework can be applied to both univariate and multivariate response data.

It is worth noting that CWMs most fundamentally differ from MoE models in their handling of the mixing proportions $\tau_g$ and in how the joint density $f\left(\mathbf{x}_i,z_{ig} = 1\right)$ is treated, either as $\Pr\negthinspace\left(z_{ig} = 1\given \mathbf{x}_i\right)=\tau_g\negthinspace\left(\mathbf{x}_i\right)$~(MoE) or $f\left(\mathbf{x}_i\given z_{ig} = 1\right)\Pr\negthinspace\left(z_{ig} = 1\right)$~(CWM). In other words, the direction of the edge between $\mathbf{X}$ and $\mathbf{Z}$ in the full MoE model in Figure \ref{Subplot:FullModel} is reversed under CWMs \citep{Ingrassia2012}. By virtue of modelling the distribution of the covariates, CWMs are also inherently less parsimonious. The~same covariate(s) can enter both parts of full MoE models, in principle. Such models~can provide a useful estimation of the conditional density of the outcome given~the covariates, but the interpretation of the clustering model and the effect of the covariates becomes more difficult in this case. Conversely, allowing different covariates enter different parts of the model further differentiates MoE models from CWMs. It is common to distinguish among the overall set of covariates between \emph{concomitant} gating network variables and \emph{explanatory} expert network variables. Thus, for clarity, $\smash{\mathbf{x}_i^{\left(G\right)}}$ and $\smash{\mathbf{x}_i^{\left(E\right)}}$ will henceforth refer, respectively, to the possibly overlapping subsets of gating and expert network covariates, such that $\smash{\mathbf{x}_i = \big\{\mathbf{x}_i^{\left(G\right)}\,\cup\,\mathbf{x}_i^{\left(E\right)}\big\}}$, with the dimensions~of the associated design matrices given by $d_G + 1$ and $d_E + 1$. Higher order terms, transformations, and interaction effects between covariates are also allowed in both networks. 

\subsection[Gaussian Parsimonious Clustering Models]{Gaussian Parsimonious Clustering Models}
\label{sec:parsimonious}
Parsimony has been considered extensively in the model-based clustering literature.~In particular, the volume of work on Gaussian and/or parsimonious mixtures has~\mbox{increased} hugely since the work of \citet{Banfield1993} and \citet{Celeux1995}. These works introduced the family of GPCMs, which are implemented in the popular \textsf{R} package \texttt{mclust} \citep{Scrucca2016}. The influence of GPCMs is clear~on many other works which obtain parsimony in the component covariance matrices;~e.g., using constrained factor-analytic structures \citep{McNicholas2008,McNicholas2010b}, the multivariate $t$-distribution and associated $t$EIGEN family \citep{Andrews2012}, and the multivariate contaminated normal distribution \citep{Punzo2016b}.  

\begin{figure}[H]
	\centering
	\includegraphics[width=0.9\textwidth, keepaspectratio]{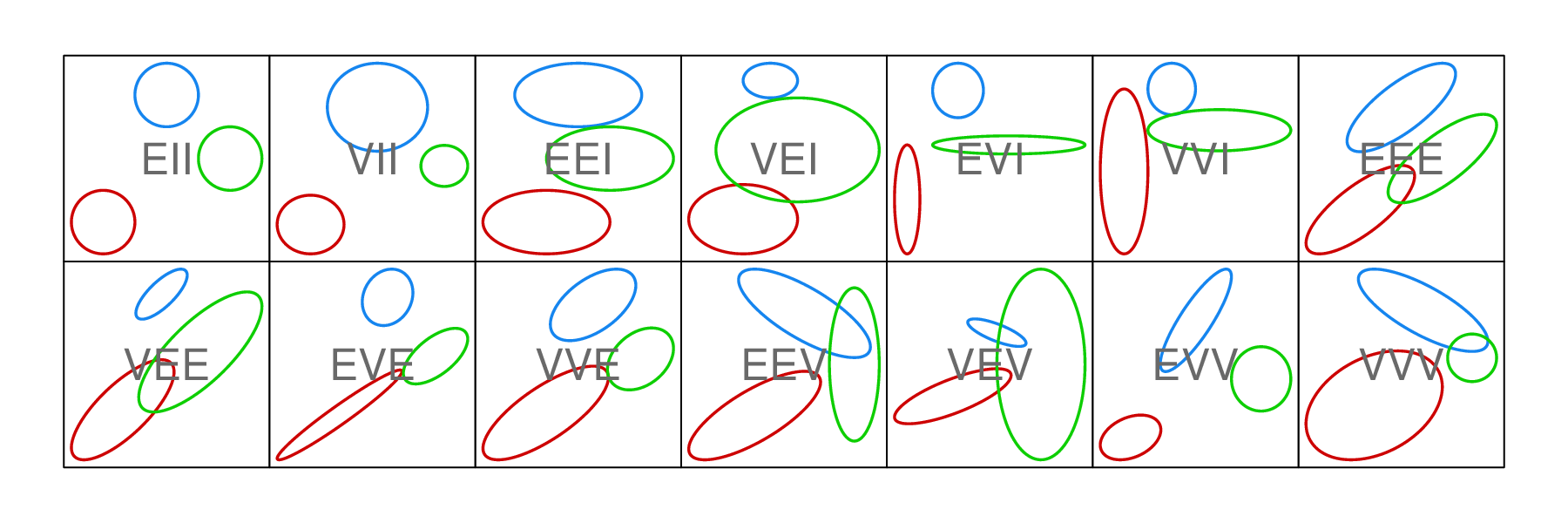}	
	\vspace{-1.125em}
	\caption{Ellipses of isodensity for each of the $14$ parsimonious eigen-decomposition covariance parameterisations for multivariate data in GPCMs, with three components in two dimensions.}
	\label{Plot:mclustellipses}
\end{figure}

Parsimonious covariance matrix parameterisations are obtained in GPCMs by means of imposing constraints on the components of an eigen-decomposition of the form $\boldsymbol{\Sigma}_g=\lambda_g\mathbf{D}_g\mathbf{A}_g\mathbf{D}_g^\top$, where $\lambda_g$ is a scalar controlling the volume, $\mathbf{A}_g$ is a diagonal matrix, with entries proportional to the eigenvalues of $\boldsymbol{\Sigma}_g$ with $\det(\mathbf{A}_g) = 1$, specifying the shape of the density contours, and $\mathbf{D}_g$ is $p\times p$ orthogonal matrix, the columns of which are the eigenvectors of $\boldsymbol{\Sigma}_g$, governing the corresponding ellipsoid's orientation. Imposing constraints reduces the number of free covariance parameters from $Gp\left(p+1\right)/2$ in the unconstrained (\textsf{VVV}) model. This is desirable when $p$ is even moderately large. Thus, GPCMs allow for intermediate component covariance matrices lying between homoscedasticity and heteroscedasticity. Table \ref{Table:MclustParams} summarises the geometric characteristics of the GPCM constraints, which are then shown in Figure \ref{Plot:mclustellipses}. 

Note for models with names ending with \textsf{I} that the number of parameters is linear in the data dimension $p$. Thus, the diagonal models are especially parsimonious and useful in $n\leq p$ settings. While there are $2$ variance parameterisations for mixtures of univariate response data, and $14$ covariance parameterisations for mixtures of multivariate response data, considering the equal mixing proportion constraint doubles the number of models available in each of these cases.
\begin{table}[H]
	\setlength{\tabcolsep}{4pt}
	\caption{Nomenclature, descriptions, and parameter counts of the parameterisations of the component covariance matrices $\boldsymbol{\Sigma}_g$ available under GPCMs --- all of which are available when there is no dependency in any way on covariates --- in increasing order of complexity. $\boldsymbol{\dagger}$ indicates availability in the first four special cases of the Gaussian MoE framework shown in Figure \ref{Plot:GraphicalRep} and the MoEClust family; $\bullet$ indicates other models available in the MoEClust family.  While all models are possible when $G=1$, they are all equivalent to one of the highlighted available models, otherwise missing entries correspond to models which are never available. The other central columns refer to $G>1$ settings.}
	\label{Table:MclustParams}
	\centering
	\scriptsize
	\vspace{-1ex}
	\begin{tabular}[pos=center]{ K{0.75cm} K{1.525cm} | K{0.8cm} K{0.75cm} K{0.75cm} | K{1.45cm} K{1cm} K{1cm} K{1.55cm} K{3cm}}
		\hline
		\centering
		Name & Model & $G=1$ & $n > p$ & $n \leq p$ & Distribution & Volume & Shape & Orientation & Covariance Parameters\\
		\specialrule{.1em}{.01em}{.01em} 
		\textsf{E} & $\sigma$ & $\boldsymbol{\dagger}$ & $\bullet$ & & (univariate) & equal & & & $1$\\
		\textsf{V} & $\sigma_g$ & & $\boldsymbol{\dagger}$ & &  (univariate) & variable & & & $G$\\
		\hline
		\textsf{EII} & $\lambda\bm{\mathcal{I}}$ & $\boldsymbol{\dagger}$ & $\bullet$ & $\bullet$ & spherical & equal & equal & --- &$1$\\
		\textsf{VII} & $\lambda_g\bm{\mathcal{I}}$ & & $\bullet$ & $\bullet$ & spherical & variable & equal & --- & $G$\\
		\textsf{EEI} & $\lambda\mathbf{A}$ & $\bullet$ &$\bullet$ & $\bullet$ & diagonal & equal & equal & axis-aligned & $p$\\
		\textsf{VEI} & $\lambda_g\mathbf{A}$ & & $\bullet$ & $\bullet$ & diagonal & variable & equal & axis-aligned & $G + (p-1)$\\
		\textsf{EVI} & $\lambda\mathbf{A}_g$ & &$\bullet$ & $\bullet$ & diagonal & equal & variable & axis-aligned & $1 + G(p-1)$\\
		\textsf{VVI} & $\lambda_g\mathbf{A}_g$ & & $\boldsymbol{\dagger}$ & $\boldsymbol{\dagger}$& diagonal & variable & variable & axis-aligned & $Gp$\\
		\textsf{EEE} & $\lambda\mathbf{D}\mathbf{A}\mathbf{D}^\top$ & $\bullet$ & $\bullet$ & & ellipsoidal & equal & equal & equal & $p(p + 1)/2$\\
		\textsf{VEE} & $\lambda_g\mathbf{D}\mathbf{A}\mathbf{D}^\top$ & & $\bullet$ & & ellipsoidal & variable & equal & equal & $G + p(p-1)/2 + (p-1)$\\
		\textsf{EVE} & $\lambda\mathbf{D}\mathbf{A}_g\mathbf{D}^\top$ &&  $\bullet$ & & ellipsoidal & equal & variable & equal & $1 + p(p-1)/2 + G(p-1)$\\
		\textsf{VVE} & $\lambda_g\mathbf{D}\mathbf{A}_g\mathbf{D}^\top$ &&  $\bullet$ & & ellipsoidal & variable & variable & equal & $G + p(p-1)/2 + G(p-1)$\\
		\textsf{EEV} & $\lambda\mathbf{D}_g\mathbf{A}\mathbf{D}_g^\top$ & & $\bullet$ & & ellipsoidal & equal & equal & variable & $1 + Gp(p-1)/2 + (p-1)$\\
		\textsf{VEV} & $\lambda_g\mathbf{D}_g\mathbf{A}\mathbf{D}_g^\top$ & & $\bullet$ & & ellipsoidal & variable & equal & variable & $G + Gp(p-1)/2 + (p-1)$\\
		\textsf{EVV} & $\lambda\mathbf{D}_g\mathbf{A}_g\mathbf{D}_g^\top$ & & $\bullet$ & & ellipsoidal & equal & variable & variable & $Gp(p+1)/2 - (G-1)$\\
		\textsf{VVV} & $\lambda_g\mathbf{D}_g\mathbf{A}_g\mathbf{D}_g^\top$ &&  $\boldsymbol{\dagger}$& & ellipsoidal & variable & variable & variable & $Gp(p+1)/2$\\
		\hline
	\end{tabular}
\end{table}

\subsection[The MoEClust Family of Models]{The MoEClust Family of Models}
\label{sec:family2}
Interest lies in bringing parsimonious covariance structures to Gaussian MoE models with network-specific subsets of covariates:\vspace{-0.75ex}
\begin{equation*}
f\big(\mathbf{y}_i \given \mathbf{x}_i\big) = \sum_{g=1}^{G}\tau_g\big(\mathbf{x}_i^{\left(G\right)}\big)\phi\Big(\mathbf{y}_i\bigm|\boldsymbol{\theta}_g\big(\mathbf{x}_i^{\left(E\right)}\big) = \left\{ \widetilde{\mathbf{x}}_i^{\left(E\right)}\boldsymbol{\gamma}_g, \boldsymbol{\Sigma}_g\right\}\Big),\vspace{-0.75ex}
\end{equation*}
where $\boldsymbol{\Sigma}_g$ can follow any of the GPCM constraints outlined in Table \ref{Table:MclustParams}. It is equivalent to say that interest lies in incorporating covariate information into the GPCM model family. Using the covariance constraints, combined with the six special cases of the MoE model described in Section \ref{sec:family}, yields the MoEClust family of models, which are capable of dealing with correlated responses and offering additional parsimony in the component densities compared to current implementations of Gaussian MoE models, by virtue of allowing the size, volume, shape, and/or orientation to be equal or unequal across components. For MoE models, every continuous covariate added to the gating and expert networks introduces $G-1$ and $Gp$ additional regression parameters, respectively. Parsimonious MoEClust models allow the increase in the number of regression parameters to be offset by the reduction in the number of covariance parameters. This can be advantageous when model selection is conducted using information criteria which include penalty terms based on parameter counts (see Section \ref{sec:selection}).

\subsection[Existing Models and Software]{Existing Models and Software}
\label{sec:software}
A number of tools for fitting MoE models are available in the \textsf{R} programming environment \citep{R2021}. These include \texttt{flexmix} \citep{Grun2007, Grun2008}, \texttt{mixtools} \citep{Benaglia2009}, and others. Tools for fitting GPCMs without covariates include \texttt{mclust} \citep{Scrucca2016} and \texttt{Rmixmod} \citep{Rmixmod}.\enlargethispage{\baselineskip}

The \texttt{flexmix} package \citep{Grun2007, Grun2008} can accommodate the full range of MoE models outlined in Section \ref{sec:family}, excluding those for which $\boldsymbol{\tau}$ is constrained to be equal, in the case of univariate $\mathbf{y}_i$, though only models with unequal variance can be fitted. The user can specify the form of the GLM and covariates (if any) to~be used in the gating and expert networks, for which the package has a similar interface~to the \texttt{glm} functions within \textsf{R}. In the case of a multivariate continuous response,~there is functionality for multivariate Gaussian component distributions though only for models without expert network covariates. Furthermore, only the \textsf{VVI} and \textsf{VVV} constraints and models with unequal mixing proportions or gating concomitants are facilitated.

For univariate data, the \texttt{mixtools} package \citep{Benaglia2009} can accommodate the expert network MoE model with equal or unequal variance; it can also accommodate the full MoE model, though only for $G=2$, with unequal variance, and with the restriction that all covariates enter both part of the model. The package allows for nonaparametric estimation of the functional form for the mixing proportions (gating networks) and the component densities (expert networks), so it offers further flexibility beyond \texttt{flexmix} in these cases. However, the multivariate models in \texttt{mixtools} use the local independence assumption, so it does not directly offer the facility to model multivariate Gaussian component densities with non-diagonal covariance matrices. Furthermore, multivariate response models in \texttt{mixtools} do not yet incorporate covariates in any way, and the equal mixing proportions constraint is not facilitated either.

The \texttt{mclust} package \citep{Scrucca2016} and \texttt{Rmixmod} package \citep{Rmixmod} can accommodate the full range of covariance constraints in Table \ref{Table:MclustParams}, and are thus examples of existing software which can fit GPCMs, but only using the standard finite mixture model (model (\subref{Subplot:MixtureModel}) in Figure \ref{Plot:GraphicalRep}) or the equal mixing proportions mixture model; i.e., they do not facilitate dependency on covariates in any way.

Another important contribution in this area is by \citet{Dang2015}. This work introduces eigen-decomposition parsimony to the MoE framework, though only for the expert network MoE model and the full MoE model. However, for the full MoE model, all covariates are assumed to enter into both parts of the model. Thus, the MoEclust model family completes the work of \citet{Dang2015} by considering all six special cases of the MoE framework, whereby different subsets of covariates can enter either, neither, or both the component densities and/or component weights, as well as models with equal mixing proportions. In addition, our unifying MoEClust framework also incorporates such parsimonious models for univariate response data.

Finally, it should be noted that eigen-decomposition parsimony has been introduced to the alternative CWM framework, in which all covariates enter the same part of the model, by \citet{Dang2017}, for the multivariate Gaussian distributions of both the response variables and the covariates, assuming only continuous covariates; see also \citet{Punzo2015} for eigen-decomposition parsimony applied to the covariates only. The \texttt{flexCWM} package \citep{Mazza2018} allows GPCM covariance structures in the distribution of the continuous covariates only, though only univariate responses are accommodated. It also allows, simultaneously or otherwise, covariates of other types, as well as omitting the distribution for the covariates entirely, leading to non-parsimonious mixtures of regressions, with or without concomitant variables.

\section[Model Fitting via EM]{Model Fitting via EM}
\label{sec:Inference}
To estimate the parameters of MoEClust models, we focus on maximum likelihood estimation using the EM algorithm \citep{Dempster1977}. This is outlined first for MoE models in Section \ref{sec:MoEInference} and then extended to MoEClust models in Section \ref{sec:MoEClustInference}. Model fitting details are described chiefly for the full MoE model only, for simplicity. A simple trick involving the residuals of the weighted linear regressions in the expert network assists fitting when using GPCM constraints. A uniform noise component to capture outlying non-Gaussian observations is added in Section \ref{sec:noise}. When gating concomitants are present, the noise component is treated in two different ways.

\subsection[Fitting MoE Models]{Fitting MoE Models}
\label{sec:MoEInference}
For the full mixture of experts model, the likelihood is of the form
\[\mathcal{L}\negthinspace\left(\boldsymbol{\beta},\boldsymbol{\gamma},\boldsymbol{\Sigma}\given\mathbf{Y},\mathbf{X}\right) = \prod_{i=1}^{n}\sum_{g=1}^{G}\tau_g\big(\mathbf{x}_i^{\left(G\right)}\big)\phi\big(\mathbf{y}_i\given\boldsymbol{\theta}_g\big(\mathbf{x}_i^{\left(E\right)}\big)\big),\]
where $\tau_g\big(\mathbf{x}_i^{\left(G\right)}\big)$ and $\boldsymbol{\theta}_g\big(\mathbf{x}_i^{\left(E\right)}\big)$ are as defined by \eqref{eq:tautheta}. The data are augmented by imputing the latent cluster membership indicator $\smash{\mathbf{z}_i=\left(z_{i1},\ldots,z_{iG}\right)^\top}$. Thus, the conditional distribution of $\big(\mathbf{y}_i,\mathbf{z}_i\given \mathbf{x}_i\big)$ is of the form
\[f\big(\mathbf{y}_i,\mathbf{z}_i\given \mathbf{x}_i\big) = \prod_{g=1}^{G}\left\lbrack\tau_g\big(\mathbf{x}_i^{\left(G\right)}\big)\phi\big(\mathbf{y}_i\given\boldsymbol{\theta}_g\big(\mathbf{x}_i^{\left(E\right)}\big)\big)\right\rbrack^{z_{ig}}\negmedspace.\]
Hence, the complete data likelihood is of the form
\[\mathcal{L}_c\negthinspace\left(\boldsymbol{\beta},\boldsymbol{\gamma},\boldsymbol{\Sigma}\given\mathbf{Y},\mathbf{X},\mathbf{Z}\right) = \prod_{i=1}^{n}\prod_{g=1}^{G}\left\lbrack\tau_g\big(\mathbf{x}_i^{\left(G\right)}\big)\phi\big(\mathbf{y}_i\given\boldsymbol{\theta}_g\big(\mathbf{x}_i^{\left(E\right)}\big)\big)\right\rbrack^{z_{ig}}\negmedspace,\]
and the complete data log-likelihood has the form
\begin{equation}
\begin{aligned}
\ell_c\negthinspace\left(\boldsymbol{\beta},\boldsymbol{\gamma},\boldsymbol{\Sigma}\given\mathbf{Y},\mathbf{X},\mathbf{Z}\right) &= \sum_{i=1}^{n}\sum_{g=1}^{G}z_{ig}\left\lbrack\log\tau_g\big(\mathbf{x}_i^{\left(G\right)}\big) + \log \phi\big(\mathbf{y}_i\given\boldsymbol{\theta}_g\big(\mathbf{x}_i^{\left(E\right)}\big)\big)\right\rbrack\\
&= \sum_{i=1}^{n}\sum_{g=1}^{G}z_{ig}\log\tau_g\big(\mathbf{x}_i^{\left(G\right)}\big) + \sum_{i=1}^{n}\sum_{g=1}^{G}z_{ig}\log \phi\big(\mathbf{y}_i\given\boldsymbol{\theta}_g\big(\mathbf{x}_i^{\left(E\right)}\big)\big).
\end{aligned}
\label{eq:completeloglike}
\end{equation}
\indent The iterative EM algorithm for MoE models follows in a similar manner to that for standard mixture models. It consists of an E-step (expectation) which replaces for each observation the~missing data $\mathbf{z}_i$ with their expected values $\smash{\widehat{\mathbf{z}}_i}$, followed by a M-step (maximisation) which maximises the expected complete data log-likelihood, computed with the estimates $\smash{\widehat{\mathbf{Z}} = \left(\widehat{\mathbf{z}}_1,\ldots,\widehat{\mathbf{z}}_n\right)}$, to provide estimates of the component weight parameters $\smash{\widehat{\tau}_g\big(\mathbf{x}_i^{\left(G\right)}\big)}$ and the component parameters $\smash{\widehat{\boldsymbol{\theta}}_g\big(\mathbf{x}_i^{\left(E\right)}\big)}$. Aitken's acceleration criterion is used to assess convergence of the non-decreasing sequence of log-likelihood estimates \citep{Boehning1994}. Parameter estimates produced on convergence achieve at least a local maximum of the likelihood function. Upon convergence, cluster memberships~are estimated via the maximum \emph{a posteriori} (MAP) classification. The E-step involves computing
\begin{equation*}
\widehat{z}_{ig}^{\left(t+1\right)} = \mathbb{E}\negthinspace\left(z_{ig}\bigm| \mathbf{y}_i, \mathbf{x}_i, \widehat{\boldsymbol{\beta}}^{\left(t\right)},\widehat{\boldsymbol{\gamma}}^{\left(t\right)},\widehat{\boldsymbol{\Sigma}}^{\left(t\right)}\right) = \frac{\widehat{\tau}_g^{\left(t\right)}\big(\mathbf{x}_i^{\left(G\right)}\big)\phi\big(\mathbf{y}_i\given\widehat{\boldsymbol{\theta}}_g^{\left(t\right)}\big(\mathbf{x}_i^{\left(E\right)}\big)\big)}{\sum_{h=1}^{G}\widehat{\tau}_h^{\left(t\right)}\big(\mathbf{x}_i^{\left(G\right)}\big)\phi\big(\mathbf{y}_i\given\widehat{\boldsymbol{\theta}}_h^{\left(t\right)}\big(\mathbf{x}_i^{\left(E\right)}\big)\big)},
\end{equation*}
where $\smash{\big\{\widehat{\boldsymbol{\beta}}^{\left(t\right)},\widehat{\boldsymbol{\gamma}}^{\left(t\right)},\widehat{\boldsymbol{\Sigma}}^{\left(t\right)}\big\}}$ are the estimates of the parameters in the gating and expert networks on the $t$-th iteration of the EM algorithm.

For the M-step, we notice that the complete data log-likelihood in \eqref{eq:completeloglike} can be considered as~a separation into the portion due to the gating network and the portion due to the expert network. Thus, the expected complete data log-likelihood \eqref{eq:separate_ll} can be maximised separately under the EM framework:
\vspace{-1.5ex}
\begin{equation}
\begin{aligned}
\mathbb{E}\negthinspace\left\lbrack\ell_c\big(\boldsymbol{\beta},\boldsymbol{\gamma},\boldsymbol{\Sigma}\given \mathbf{Y}, \mathbf{X},\mathbf{Z},\widehat{\boldsymbol{\beta}}^{\left(t\right)},\widehat{\boldsymbol{\gamma}}^{\left(t\right)},\widehat{\boldsymbol{\Sigma}}^{\left(t\right)} \big)\right\rbrack&=\sum_{i=1}^{n}\sum_{g=1}^{G}\widehat{z}^{\left(t + 1\right)}_{ig}\log\tau_g\big(\mathbf{x}_i^{\left(G\right)}\big)\\[-0.5ex] &+\sum_{i=1}^{n}\sum_{g=1}^{G}\widehat{z}^{\left(t + 1\right)}_{ig}\log\phi\big(\mathbf{y}_i\given\boldsymbol{\theta}_g\big(\mathbf{x}_i^{\left(E\right)}\big)\big).\label{eq:separate_ll}\vspace{-1ex}
\end{aligned}
\vspace{-1ex}
\end{equation}
\noindent The first term is of the same form as a MLR model, here written with component $1$ as the baseline reference level, for identifiability reasons:\vspace{-0.75ex}
\begin{equation*}\log \frac{\tau_g\big(\mathbf{x}_i^{\left(G\right)}\big)}{\tau_1\big(\mathbf{x}_i^{\left(G\right)}\big)}  = \log\frac{\Pr\negthinspace\big(\widehat{z}^{\left(t + 1\right)}_{ig}=1\big)}{\Pr\negthinspace\big(\widehat{z}^{\left(t + 1\right)}_{i1}=1\big)} = \widetilde{\mathbf{x}}_i^{\left(G\right)}\boldsymbol{\beta}_g~\forall~g\geq 2,~\mbox{where}~\boldsymbol{\beta}_1 = \left(0,\ldots,0\right)^\top\negmedspace.\vspace{-0.75ex}
\end{equation*}
Thus, methods for fitting such models can be used to estimate the parameters of the gating network. However, due to the nonlinear numerical optimisation involved in estimating MLR coefficients, for which no closed-form updates are available, we caution that this step merely increases the expectation of this portion of \eqref{eq:separate_ll} at each iteration, without explicitly maximising it. Nevertheless,~the monotonicity of the EM is preserved and the procedure is still guaranteed to converge \citep{Dempster1977}. 

The second term is of the same form as fitting $G$ separate weighted multivariate linear regressions, and thus methods for fitting such models can be used to estimate the expert network parameters. Note that these are multivariate in the sense of a multivariate outcome $\smash{\mathbf{y}_i}$; the associated design matrix having $\smash{d_E + 1}$ columns means these regressions are possibly also multivariate in terms of the explanatory variables. Thus, fitting MoE models is straightforward in principle.

\subsection[Fitting MoEClust Models]{Fitting MoEClust Models}
\label{sec:MoEClustInference}
Maximising the second term in \eqref{eq:separate_ll}, corresponding to the expert network, gives rise to the following expression\vspace{-1.5ex}
\begin{equation}
\begin{split}
-\frac{1}{2}\Big(&p\log 2\pi + \sum_{i=1}^{n}\sum_{g=1}^{G}\widehat{z}_{ig}^{\left(t+1\right)}\log \lvert\boldsymbol{\Sigma}_g\rvert + \Big. \\[-1.25ex]
\Big.& \sum_{i=1}^n\sum_{g=1}^G \widehat{z}_{ig}^{\left(t+1\right)}\big(\mathbf{y}_i -  \widetilde{\mathbf{x}}_i^{\left(E\right)}\boldsymbol{\gamma}_g\big)^\top\boldsymbol{\Sigma}_g^{-1}\big(\mathbf{y}_i -  \widetilde{\mathbf{x}}_i^{\left(E\right)}\boldsymbol{\gamma}_g\big)\Big).\\[-1ex]
\end{split}\label{eq:expertll}
\end{equation}
When the same set of regressors are used for each dependent variable, as is always the case for MoEClust models, or when $\smash{\boldsymbol{\Sigma}_g}$ is diagonal, it can be shown that $\smash{\boldsymbol{\gamma}_g}$ does not depend on $\smash{\boldsymbol{\Sigma}_g}$, much like a Seemingly Unrelated Regression model (SUR; \citealp{Zellner1962}). We first estimate $\smash{\widehat{\boldsymbol{\gamma}}_g}$ and then $\smash{\widehat{\boldsymbol{\Sigma}}_g}$. Fitting $G$ separate multivariate regressions (weighted by $\smash{\widehat{z}_{ig}}$), yields $G$ sets of $n \times p$ SUR residuals $\smash{\widehat{\mathbf{r}}_{ig} = \mathbf{y}_i -  \widetilde{\mathbf{x}}_i^{\left(E\right)}\widehat{\boldsymbol{\gamma}}_g}$ which, crucially, satisfy $\smash{\sum_{i=1}^{n}\widehat{z}_{ig}\widehat{\mathbf{r}}_{ig}=0}$. Thus, maximising \eqref{eq:expertll} is equivalent to minimising\vspace{-1ex}\enlargethispage{\baselineskip}
\begin{equation}
\label{eq:Mstep}
\sum_{i=1}^{n}\sum_{g=1}^{G}\widehat{z}_{ig}^{\left(t+1\right)}\log\lvert\boldsymbol{\Sigma}_g\rvert +\sum_{i=1}^{n}\sum_{g=1}^{G}\widehat{z}_{ig}^{\left(t+1\right)}\widehat{\mathbf{r}}_{ig}^\top\boldsymbol{\Sigma}_g^{-1}\widehat{\mathbf{r}}_{ig}^{\phantom{\top}}\,,\vspace{-1ex}
\end{equation}
which is of the same form as the criterion used in the M-step of a standard \mbox{Gaussian} finite mixture model with component covariance matrices $\smash{\widehat{\boldsymbol{\Sigma}}}$, component means equal~to zero, and new augmented data set $\smash{\widehat{\mathbf{R}}}$. Thus, when estimating the component covariance matrices via \eqref{eq:Mstep}, the same M-step function as used within \texttt{mclust} can be applied to augmented data, constructed so that each observation is represented as follows:

\begin{enumerate}[noitemsep]
	\mathleft
	\begin{minipage}[t]{0.49\linewidth}
		\item \small{Stack the $G$ sets of SUR residuals\\ into the $\left(n \times G\right) \times p$ matrix $\smash{\widehat{\mathbf{R}}}$:} \vspace{-1.25ex}
		\[\widehat{\mathbf{R}} = {\scriptsize\begin{bmatrix}
		\widehat{r}_{111} & \widehat{r}_{112} & \dots  & \widehat{r}_{11p} \\
		\widehat{r}_{211} & \widehat{r}_{212} & \dots  & \widehat{r}_{21p} \\
		\vdots  & \vdots  & \ddots & \vdots  \\
		\widehat{r}_{n11} & \widehat{r}_{n12} & \dots  & \widehat{r}_{n1p} \\
		\hline
		\widehat{r}_{121} & \widehat{r}_{122} & \dots  & \widehat{r}_{12p} \\
		\widehat{r}_{221} & \widehat{r}_{222} & \dots  & \widehat{r}_{22p} \\
		\vdots  & \vdots  & \ddots & \vdots  \\
		\widehat{r}_{n21} & \widehat{r}_{n22} & \dots  & \widehat{r}_{n2p} \\
		\hline
		\vdots  & \vdots  & \ddots & \vdots  \\
		\hline
		\widehat{r}_{1G1} & \widehat{r}_{1G2} & \dots  & \widehat{r}_{1Gp} \\
		\widehat{r}_{2G1} & \widehat{r}_{2G2} & \dots  & \widehat{r}_{2Gp} \\
		\vdots  & \vdots  & \ddots & \vdots  \\
		\widehat{r}_{nG1} & \widehat{r}_{nG2} & \dots  & \widehat{r}_{nGp} \\
		\end{bmatrix}}\]
	\end{minipage}\hfill%
	\begin{minipage}[t]{0.49\linewidth}
		\item \small{Create the $\left(n \times G\right) \times G$ block-diagonal\\matrix $\smash{\widehat{\boldsymbol{\zeta}}}$ from the columns of $\widehat{\mathbf{Z}}$:}\vspace{-1.25ex}
		\[\widehat{\boldsymbol{\zeta}} = {\scriptsize\begin{bmatrix}
		\widehat{z}_{11} & 0 & \dots  & 0 \\
		\widehat{z}_{21} & 0 & \dots  & 0 \\
		\vdots  & \vdots & \ddots & \vdots  \\
		\widehat{z}_{n1} & 0 & \dots  & 0 \\
		\hline
		0 & \widehat{z}_{12} & \dots  & 0 \\
		0 & \widehat{z}_{22} & \dots  & 0 \\
		\vdots  & \vdots & \ddots & \vdots  \\
		0 & \widehat{z}_{n2} & \dots  & 0 \\
		\hline
		\vdots  & \vdots & \ddots & \vdots  \\
		\hline
		0 & 0 & \dots  & \widehat{z}_{1G} \\
		0 & 0 & \dots  & \widehat{z}_{2G} \\
		\vdots  & \vdots & \ddots & \vdots  \\
		0 & 0 & \dots  & \widehat{z}_{nG} \\
		\end{bmatrix}}\]
	\end{minipage}
	\mathcenter
\end{enumerate}
\vspace{-1ex}
Structuring the model in this manner allows GPCM covariance structures to be easily imposed on Gaussian MoE models with gating and/or expert network covariates. In the end, the M-step involves three sub-steps,~each using the current estimate of $\smash{\widehat{\mathbf{Z}}}$: i) estimating the gating network parameters $\smash{\widehat{\boldsymbol{\beta}}}_g$ and hence the component weights $\smash{\widehat{\tau}_g\big(\mathbf{x}_i^{\left(G\right)}\big)}$ via MLR, ii) estimating the expert network parameters $\smash{\widehat{\boldsymbol{\gamma}}_g}$ and hence the component-specific means via weighted multivariate multiple linear regression, and iii) estimating the constrained component covariance matrices $\smash{\widehat{\boldsymbol{\Sigma}}_g}$ using the augmented data set comprised of SUR residuals, as outlined above. 

In the absence of covariates in the gating and/or expert networks, under the special cases outlined in Section \ref{sec:family}, their respective contribution to \eqref{eq:separate_ll} is maximised as~per the corresponding term in a standard GPCM. In other words, the gating and expert networks without covariates can be seen as regressions with only an intercept term. Thus, the augmented data structure is not required~when there are no expert covariates and the formula for estimating $\boldsymbol{\tau}$ in the absence of concomitant variables is $\widehat{\tau}_g = n^{-1}\sum_{i=1}^n \widehat{z}_{ig}$, rather than \eqref{eq:tautheta}. As described in Section \ref{sec:family}, it is sometimes useful~to expand the model family further by considering more parsimonious alternatives to the special cases of models \ref{eq:MoEmix} and \ref{eq:MoEexp} in Figure \ref{Plot:GraphicalRep}, where gating covariates are omitted,~by constraining the mixing proportions to be equal and fixed, i.e. $\tau_g=\nicefrac{1}{G}\:\:\forall\:\:g$. Similarly, removing the corresponding regression intercept(s) from the part(s) of the model where covariates enter can yield further parsimony in appropriate settings, e.g.~when there~are strong \emph{a priori} physical reasons for believing $\mathbb{E}\left(\mathbf{Y}\given  \mathbf{X}^{\left(E\right)} = \mathbf{0}\right)=\mathbf{0}$ \citep{Eisenhauer2003}.\vspace{-0.5ex}

\subsection[Adding a Noise Component]{Adding a Noise Component}
\label{sec:noise}
For models with expert network covariates, and/or when the volume and/or shape~\mbox{differ} across components, the mixture likelihood is unbounded. We restrict our interest only to solutions for which the log-likelihood at convergence is finite. As per the~\texttt{eps} argument to the \texttt{mclust} \textsf{R} package's \texttt{emControl} function \citep{Scrucca2016}, we monitor the conditioning of the covariances and add a tolerance parameter (set to the relative machine precision, i.e. \texttt{2.220446e-16} on IEEE compliant machines) to the M-step estimation of the component covariances to control termination of the EM algorithm on the basis of small eigenvalues. For models with unconstrained $\boldsymbol{\Sigma}_g$, each cluster must contain at least $p + 1$ units to avoid computational singularity. Thus,~in practice, such spurious solutions with infinite likelihood occur especially for higher~$G$ values, whereby either solutions with empty components reduce to ones with fewer components, or uninteresting solutions with degenerate components containing too few units or even singletons are found. Sensible initial allocations (see Section \ref{sec:init})~and/or the equal mixing proportion constraint, which help avoid empty or otherwise poorly populated clusters, can help to alleviate this problem. \citet{Garcia2018} offer an excellent discussion of the notions of spurious solutions and degenerate components.

Further extending MoEClust models via the inclusion of an additional uniform noise component can also help in addressing these issues, by capturing outlying observations which do not fit the prevailing pattern of Gaussian clusters and thus would otherwise be assigned to (possibly many) small clusters. In particular, the noise component for encompassing clusters with non-Gaussian distributions is here distributed as~a homogeneous spatial Poisson process, as per \citet{Banfield1993}. Such a noise component can be included regardless of where covariates (if any) enter, and regardless of the GPCM constraints employed
. Model-fitting via the EM algorithm is not greatly complicated by the addition of a noise component, though it is required to estimate $V$, the hypervolume of the region from which the response data have been drawn, or to consider $V$ as an independent tuning parameter as per \citet{Hennig2008}, especially if $n\leq p$. For univariate responses $V$ is given by the range of $y_1,\ldots,y_n$. For multivariate data, $V$ can be estimated by the hypervolume of the convex hull, ellipsoid hull, or smallest hyperrectangle enclosing the data. We focus on the latter method.

For initialisation, a column in which each entry is $\tau_0$ (the guess of the prior probability that obser\-vations are noise) is appended to the starting $\mathbf{Z}$ matrix, with other columns corresponding to non-noise components then multiplied by $1-\tau_0$. The initial $\tau_0$ should not be too high; it is set to $0.1$ here. For models with a noise component and no gating concomitants, the mixing proportions can be, as before, either constrained or unconstrained. In the latter case, we estimate $\tau_0$ and then constrain the remaining proportions. We add the extension that concomitants, when present, are allowed~to affect \eqref{eq:noisegate} or not affect \eqref{eq:noisenogate} the mixing proportion of the noise component. Henceforth,~for clarity, we refer to these settings as the gated noise (NG) and non-gated noise (NGN) settings, respectively. The NGN setting assumes $\tau_0$ is constant across observations and covariate patterns. It is thus the more parsimonious model; it requires only $1$ extra gating network parameter, rather than $d_G + 1$ under the GN setting, relative to models without a noise component, though it is only defined for~$G\geq2$.
\begin{align}
\mbox{GN}\colon\quad f\big(\mathbf{y}_i \given \mathbf{x}_i\big) &= \sum_{g=1}^{G}\tau_g\big(\mathbf{x}_i^{\left(G\right)}\big)\phi\Big(\mathbf{y}_i\bigm|\boldsymbol{\theta}_g\big(\mathbf{x}_i^{\left(E\right)}\big) = \left\{\widetilde{\mathbf{x}}_i^{\left(E\right)}\boldsymbol{\gamma}_g, \boldsymbol{\Sigma}_g\right\}\Big) + \frac{\tau_0\big(\mathbf{x}_i^{\left(G\right)}\big)}{V}.\label{eq:noisegate}\\
\mbox{NGN}\colon\quad f\big(\mathbf{y}_i \given \mathbf{x}_i\big) &= \sum_{g=1}^{G}\tau_g\big(\mathbf{x}_i^{\left(G\right)}\big)\phi\Big(\mathbf{y}_i\bigm|\boldsymbol{\theta}_g\big(\mathbf{x}_i^{\left(E\right)}\big) = \left\{\widetilde{\mathbf{x}}_i^{\left(E\right)}\boldsymbol{\gamma}_g, \boldsymbol{\Sigma}_g\right\}\Big) + \frac{\tau_0}{V}.\label{eq:noisenogate}
\end{align}
\vspace{-2em}

\section[Practical Issues]{Practical Issues}
\label{sec:Performance}
In this section, factors affecting the performance of MoEClust models are discussed; namely, the necessity of a good initial partition to prevent the EM algorithm from converging to a suboptimal local maximum (Section \ref{sec:init}), and the necessity of model selection with regard to where and what covariates (if any) enter the model to yield further parsimony by reducing the number of gating and/or expert network regression parameters (Section \ref{sec:selection}). Novel strategies for dealing with both issues are proposed.

\subsection[EM Initialisation]{EM Initialisation}
\label{sec:init}
With regards to initialisation of the EM algorithm for $G>1$ MoEClust models, model-based agglomerative hierarchical clustering and quantile-based clustering have been found to be suitable for multivariate and univariate data, respectively. Both \texttt{flexmix} and \texttt{mixtools} randomly initialise the allocations, despite the obvious computational drawback of the need to run the EM algorithm from multiple random starting points. However, when explanatory variables $\smash{\mathbf{x}_i^{\left(E\right)}}$ enter the expert network, it is useful to use them to augment the initialisation strategy with extra steps.  Algorithm \ref{Alg:InitEM} outlines the proposed initialisation strategy, which is similar to that of \citet{Ning2008} but modified to account for the multivariate nature of the response. It takes the initial partition of the data (whether obtained by hierarchical clustering, random initialisation, or some other method) and iteratively reallocates observations in such a way that each subset can be well-modelled by a single expert.

When using a deterministic approach to obtain the starting partition for Algorithm \ref{Alg:InitEM}, initialisation can be further improved by considering information in the expert network covariates to find a good clustering of the joint distribution of $\smash{\big(\mathbf{y}_i, \mathbf{x}_i^{\left(E\right)}\big)}$. When $\smash{\mathbf{x}_i^{\left(E\right)}}$ includes categorical or ordinal covariates, the model-based approach to clustering mixed-type data of \citet{McParland2016} can be employed at this stage, though this is not considered further here.

\bigskip
\begin{algorithm}[H]
	\caption{\small{Iterative reallocation initialisation with expert network covariates}}
	\label{Alg:InitEM}
	\footnotesize
	\setcounter{AlgoLine}{-1}
	Concatenate the response data and expert network covariates into a matrix.\\[1ex]
	Obtain some non-overlapping hard starting partition $\boldsymbol{\Omega}_1, \boldsymbol{\Omega}_2, \ldots, \boldsymbol{\Omega}_G$.\\[1ex]
	Estimate the expert network regression $\eta_g\negthinspace\left(\boldsymbol{\gamma}_g, \cdot\right)$ on every subset  $\left\{\boldsymbol{\Omega}_g\right\}_{g=1}^{G}$.\\[1ex]
	Compute the fitted values $\widehat{\mathbf{y}}_{ig} = \eta_g\big(\widehat{\boldsymbol{\gamma}}_g, \mathbf{x}_i^{\left(E\right)}\big)\:\:\forall\:\:(i, g)$ and hence the residuals $\widehat{\mathbf{r}}_{ig}=\mathbf{y}_i -  \widehat{\mathbf{y}}_{ig}$.\\[1ex]
	Compute $\widehat{\boldsymbol{\Psi}}_g = \textrm{Cov}\big(\widehat{\mathbf{R}}_g\big)=\frac{1}{n-d_E-1}\widehat{\mathbf{R}}_g^\top\widehat{\mathbf{R}}_g^{\phantom{\top}}\:\forall\:\:g$.\\[1ex]
	Compute the squared Mahalanobis distance $\widehat{M}_{ig} = \mathrm{d}^2_{\mathrm{M}}\big(\mathbf{y}_i, \widehat{\mathbf{y}}_{ig}\big) =\widehat{\mathbf{r}}_{ig}^\top\widehat{\boldsymbol{\Psi}}_g^{-1}\widehat{\mathbf{r}}_{ig}^{\phantom{\top}}\:\:\forall\:\:(i,g)$.\\[1ex]
	Let $k_i = \argmin_{h\,\in\,\left\{1,\ldots,G\right\}} \big(\widehat{M}_{ih}\big)$ and reassign observation $i$ to subset $\boldsymbol{\Omega}_{k_i}$.\\[1ex]
	Repeat Steps $2\mbox{--}6$ until convergence is achieved, i.e. until the partition ceases to change.
\end{algorithm}
\bigskip

If at any stage a level is dropped from a categorical variable in subset $\boldsymbol{\Omega}_g$ the variable itself is dropped from the corresponding regressor for the observations with missing levels. Convergence of the algorithm is guaranteed and the additional computational burden incurred is negligible. By using the Mahalanobis distance metric \citep{Mahalanobis1936}, each observation is assigned to the cluster corresponding to the Gaussian ellipsoid to which it is closest. This has the added advantage of potentially speeding up the running of the EM algorithm. The estimates of $\widehat{\boldsymbol{\gamma}}_g$ at convergence are used as starting values for the expert network. The gating network is initialised~by considering the partition itself at convergence as a discrete approximation of the gates. 

While convergence is monitored via the partition itself, Algorithm \ref{Alg:InitEM} implicitly finds the hard partition which minimises the total intra-component regression error criterion
\begin{equation}
\sum_{g=1}^G\minsub{\left\{\eta_g,\, \boldsymbol{\gamma}_g\right\}}\negthinspace\bigg(\sum_{i\,\in\,\boldsymbol{\Omega}_g}\mathrm{d}^2_{\mathrm{M}}\Big(\mathbf{y}_i, \eta_g\big(\boldsymbol{\gamma}_g, \mathbf{x}_i^{\left(E\right)}\big)\Big)\bigg).\label{eq:InitCrit}
\end{equation}
However, there are a few small caveats. Firstly, it suffices to use the Euclidean distance in place of the Mahalanobis distance for applications to univariate response data. Secondly, the Moore-Penrose pseudo-inverse \citep{Moore1920} or $p$-dimensional identity matrix $\bm{\mathcal{I}}_p$ is used in place of $\widehat{\boldsymbol{\Psi}}_g^{-1}$ when $n \leq p$. Finally, we note that Algorithm \ref{Alg:InitEM} applies only to the non-noise components; in the presence of a noise component, the $\widehat{\mathbf{Z}}$ matrix outputted by the algorithm at convergence is modified in the usual way.

\begin{figure}[H]
	\centering
	\begin{subfigure}[t]{.333\linewidth}
		\centering
		\captionsetup{format=hang}
		\includegraphics[width=\textwidth, keepaspectratio]{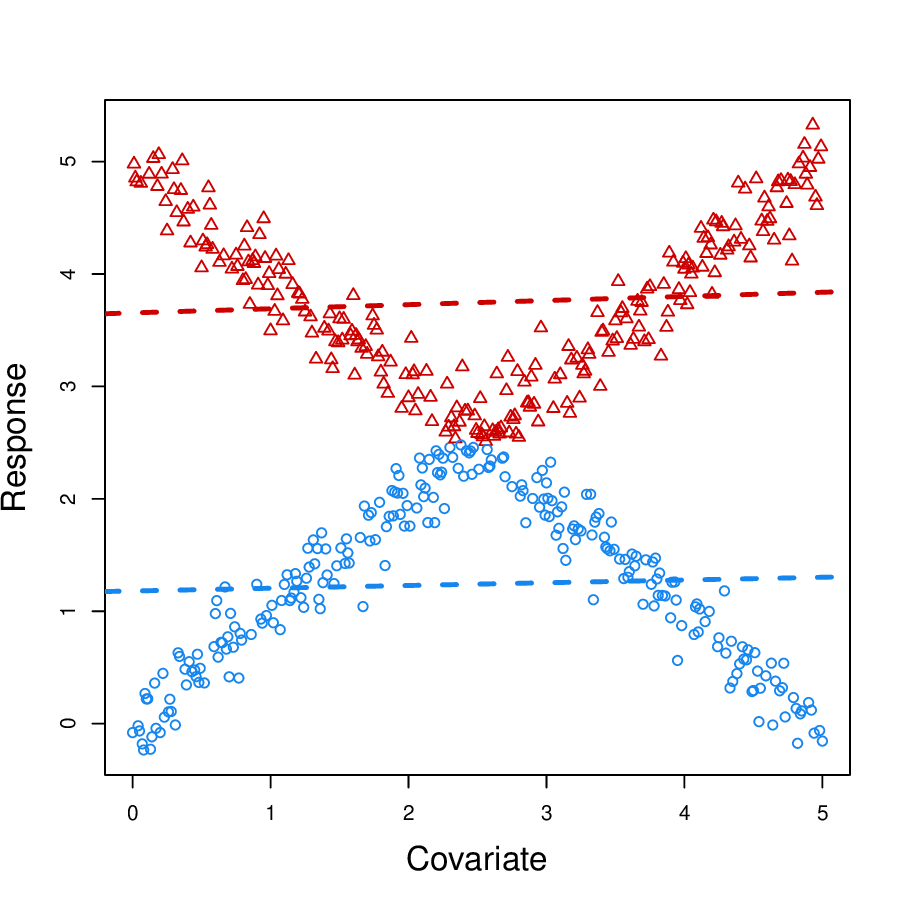}	
		\caption{Hierarchical clustering of the\\response variables only.}
		\label{Subplot:BadInit}
	\end{subfigure}\hfill%
	\begin{subfigure}[t]{.333\linewidth}
		\centering
		\includegraphics[width=\textwidth, keepaspectratio]{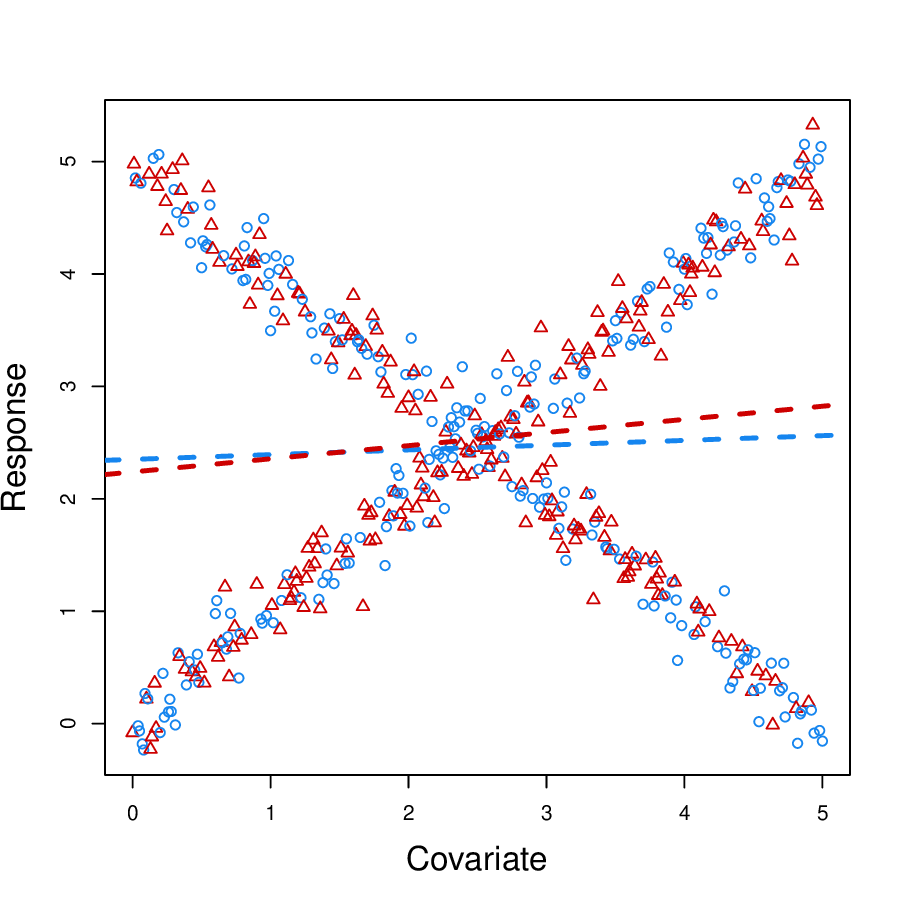}	
		\caption{Random allocation.}
		\label{Subplot:RandomInit}
	\end{subfigure}\hfill%
	\begin{subfigure}[t]{.333\linewidth}
		\centering
		\includegraphics[width=\textwidth, keepaspectratio]{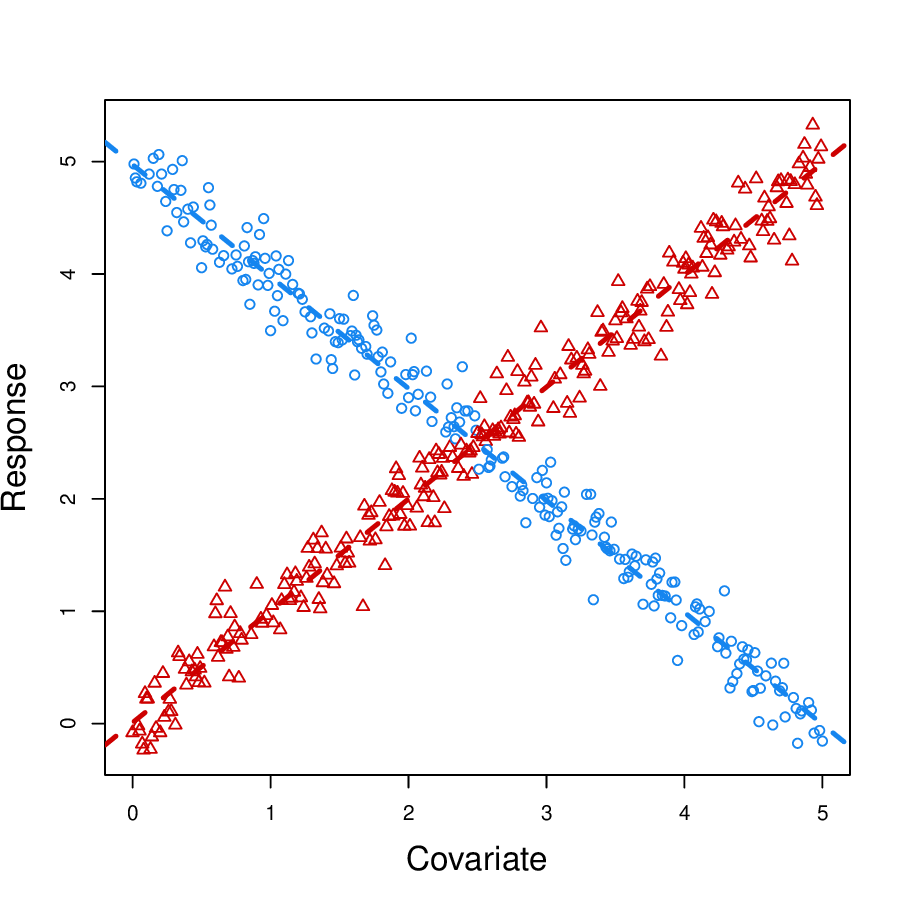}	
		\caption{Partition obtained via Algorithm \ref{Alg:InitEM}.}
		\label{Subplot:MahalaInit}
	\end{subfigure}%
	\caption{Initial $2$-component hard partitions on univariate data clearly arising from a mixture of two linear regressions, obtained using (\subref{Subplot:BadInit}) agglomerative hierarchical clustering, (\subref{Subplot:RandomInit}) random allocation,~and (\subref{Subplot:MahalaInit}) Algorithm \ref{Alg:InitEM} applied to the initialisation in (\subref{Subplot:RandomInit}) upon convergence after $6$ iterations, demonstrating~the improvement achieved by~incorporating expert network covariates into the initialisation strategy.~Allo\-cations are distinguished by blue circles and red triangles. Corresponding fitted lines are also shown.}\label{Plot:ExpertInit}
\end{figure}

Figure \ref{Plot:ExpertInit} illustrates the necessity of this procedure using a toy data set, with a single continuous covariate and a univariate response clearly arising from a mixture~of two linear regressions, which otherwise would not be discerned without including the covariate in the initialisation routine via Algorithm \ref{Alg:InitEM}. A further demonstration of the utility of this strategy is shown in Appendix\,\ref{Appendix:CO2init}. Similar to the EM algorithm's susceptibility to local maxima, a limitation of our initialisation strategy is that the result~at convergence may represent a suboptimal local minimum. However, the problem is transferred from the difficult task of initialising the EM algorithm to initialising Algorithm \ref{Alg:InitEM}. Thus, it is feasible to repeat the algorithm with many different partitions and choose the best result, in the sense of minimising the criterion in \eqref{eq:InitCrit}, to initialise one run of the EM algorithm, since Algorithm \ref{Alg:InitEM} converges very quickly, requires much less computational effort than the EM algorithm itself, and generally reduces the number~of required EM iterations. However, we caution against using the total intra-component regression error criterion to guide the inclusion of expert network covariates.\vspace{-1.5em}
\vspace{-1.5em}

\subsection[Model Selection]{Model Selection}
\label{sec:selection}
\vspace{-0.5ex}
Whether a variable should be considered as a covariate or part of the response is usually clear from the context of the data being clustered. However, within the suite of MoE models outlined in Section \ref{sec:family}, it is natural to question which covariates, if any, are~to be included, and if so in which part(s) of the MoE model. Unless the manner in which covariates enter is guided by the question of interest in the application under study, this is a challenging problem as the space of MoE models is potentially very large once variable selection for the covariates entering the gating and expert networks~is considered. Thus, only models where covariates enter all mixture components (and, in doing, affect the means of all $p$ response variables) and/or all component weights in a linear manner are typically considered in practice in order~to restrict the size of the model search space. However, even within this reduced space, there are $2^r$ models to consider when $G=1$ and $2^{2r}$ models to consider other\-wise. Thus, the model space increases further if the number of components $G$ is unknown, as is often the case in unsupervised clustering tasks.

Model comparison for the MoEClust family is even more challenging, especially for multivariate response data for which there are potentially $14$ different GPCM covariance constraints to consider for models with $G \geq 2$ and $3$ otherwise. When $p=1$, there are $2$ covariance constraints to consider when $G \geq 2$ and $1$ otherwise. Considering constraints on the mixing proportions further increases the model search space. However, model selection can still be implemented in a similar manner to other model-based clustering methods: the Bayesian Information Criterion (BIC; \citealp{Schwarz1978}) and Integrated Completed Likelihood (ICL; \citealp{Biernacki2000}) have been shown to give suitable model selection criteria, both for the number of component densities (and thus clusters) required and for selecting covariates to include in the model. The number of free parameters in the penalty term for these criteria of course depends on the included gating and expert network covariates and the GPCM constraints employed.

For MoEClust models involving mixtures of GLMs, stepwise variable selection app\-roaches can be used to find the optimal covariates for inclusion in either the multino\-mial logistic regression (gating network) or the weighted linear regression (expert network). Indeed, more parsimony can be achieved using variable selection, as there are a total of $\smash{G\left(d_G + 1\right) + Gp\left(d_E + 1\right)}$ intercept and regression coefficients to estimate for a $G > 1$ full MoE model. However, the selected covariates may only be optimal for the given $G$ and the given set of GPCM covariance matrix constraints. MoEClust models also allow for covariates entering only one part of the model. Thus, we propose a greedy stepwise search whereby each step could involve adding/removing a component or adding/removing a single covariate in either the gating or expert networks. We adopt a forward search, starting from~a $G=1$ model, as backward selection can be particularly cumbersome when $r$ is large. In the considered applications, it sufficed to consider only additions (of components and covariates) rather than additions and removals in the sense that the same final model was obtained despite fewer models being evaluated over the course of the search. Hence, the recommended forward search algorithm proceeds as follows: 

\bigskip
\begin{algorithm}[H]
	\caption{\small{Greedy forward stepwise search for MoEClust models}}
	\label{Alg:Stepwise}
	\footnotesize
	Choose the best $G=1$ model with no covariates among all allowable model types.\\[1ex]
    Either:\hspace{100em} $\bullet$~increase $G$ by $1$, \hspace{100em} $\bullet$~add an explanatory variable to the expert network, \hspace{100em} $\bullet$~add a concomitant variable to the gating network (only when $G\geq2$).\\[1ex]
    For every action in Step $2$, consider the full range of allowable GPCM constraints.\\[1ex]
    Accept the change which yields the best improvement in terms of BIC or ICL.\\[1ex]
    Repeat Steps $2\mbox{--}4$ until there is no further improvement in the selection criterion.
\end{algorithm}
\bigskip

While one could consider changing the GPCM constraints as another potential action in Step $2$ of Algorithm \ref{Alg:Stepwise}, our experience suggests that increasing $G$ or adding covariates (especially in the expert network) can radically alter the covariance structure. Thus, we advise changing the GPCM constraints simultaneously and identifying the optimum action by first finding the optimum constraints for each action. While this is more computationally intensive than altering the GPCM constraints as a step in itself, this makes the search less likely to miss optimal models as it traverses the model space. Notably, avoiding the duplication of the EM initialisation routine for certain steps involving only the addition of a gating network covariate speeds up the algorithm somewhat. See Appendix \ref{Appendix:CO2code} for an example of how to conduct such a stepwise search using code from the \texttt{MoEClust} \textsf{R} package \citep{MoEClustR2021}. 

In certain special instances, some extra steps can be considered. When there are no gating network concomitants, a choice can be made, for each action, between fitted models with equal or unequal mixing proportions. We distinguish between $G$-component models without a noise component and models with $G-1$ Gaussian components plus an additional noise component. Thus, we recommend treating models~with a noise component differently, by running a stepwise search for models excluding the possibility of a noise component, running a separate stepwise search starting from a $G=0$ noise-only model, and ultimately choosing between the optimal models with and without a noise component identified by each search. In the presence of a noise component, one can also fit models under the GN and NGN settings, given by \eqref{eq:noisegate} and \eqref{eq:noisenogate} respectively, when evaluating every action involving models with gating concomitants.

When $r$ is not so prohibitively large as to render an exhaustive search infeasible, \citet{Gormley2010} demonstrate how model selection criteria such as the BIC can be employed to choose the appropriate number of components and guide the inclusion of covariates across the six special cases of the MoE model described in Section \ref{sec:family}. Adapting this approach to MoEClust models where GPCM constraints must also be chosen requires fixing the covariates to be included in the component weights and densities and finding the $G$ value and GPCM covariance structure which together optimise some criterion. Different fits with different combinations of covariates are then compared according to the same criterion. However, due to the highlighted computational difficulties when $r$ is large, Algorithm \ref{Alg:Stepwise} remains the recommended approach.

\section[Results]{Results}
\label{sec:Results}
The clustering performance of the MoEClust models is illustrated by application to two well-known data sets: univariate CO\textsubscript{2} data (Section \ref{sec:CO2}) and multivariate data from the Australian Institute of Sports (Section \ref{sec:AIS}). Additional results are provided for each data set in the Appendices. In particular, code examples (Appendix \ref{Appendix:CO2code}) and details of the initialisation (Appendix \ref{Appendix:CO2init}) for the CO\textsubscript{2} data and results of the stepwise search (Appendix \ref{Appendix:AISstep}) for the AIS data are given.

Hereafter, any mention of methods for initialising the allocations, when covariates enter the expert network, refers to finding a single initial partition for Algorithm \ref{Alg:InitEM}. The BIC and the stepwise search strategy out\-lined in Algorithm \ref{Alg:Stepwise} were used to find the optimal number of components, choose the covariance type, and select the best subset of covariates, as well as where to put them. Results of exhaustive searches are also provided for demonstrative purposes. All results were obtained using the \textsf{R} package \texttt{MoEClust} \citep{MoEClustR2021}.

\subsection[CO\textsubscript{2} Data]{CO\textsubscript{2} Data}
\label{sec:CO2}
As a univariate example of an application of MoEClust, data on the CO\textsubscript{2} emissions of $n=28$ countries in the year 1996 \citep{Hurn2003} are clustered, with Gaussian component densities. Studying the relationship between CO\textsubscript{2} and the covariate Gross National Product (GNP), both measured \emph{per capita}, is of interest. As consideration is only being given to inclusion/exclusion of a single covariate in the gating and/or expert networks, an exhaustive search is feasible. A range of models ($G\in\{1,\ldots,9\}$)~are fitted, with either the equal (\textsf{E}) or unequal variance (\textsf{V}) models from Table \ref{Table:MclustParams}. Quantile-based clustering of the CO\textsubscript{2} values is used to initialise Algorithm \ref{Alg:InitEM} when the expert network excludes GNP, otherwise hierarchical clustering of both CO\textsubscript{2} and GNP is used.

\begin{figure}[H]
	\centering
	\begin{subfigure}[t]{.49\linewidth}
		\centering
		\includegraphics[width=\textwidth, keepaspectratio]{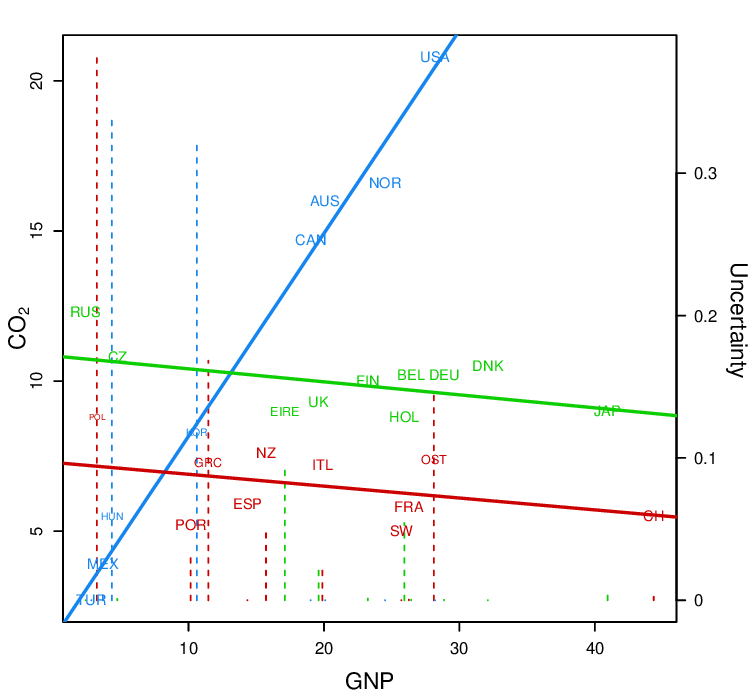}	
		\caption{Fitted lines of the expert network GLMs. Text label size is proportional to a country's probability of belonging to its assigned cluster. Clustering uncertainty is also indicated by dotted vertical bars relating to the second y-axis. Colours correspond to the MAP classification.}
		\label{Plot:CO2Expert}
	\end{subfigure}\hfill%
	\begin{subfigure}[t]{.49\linewidth}
		\centering
		\includegraphics[width=\textwidth, keepaspectratio]{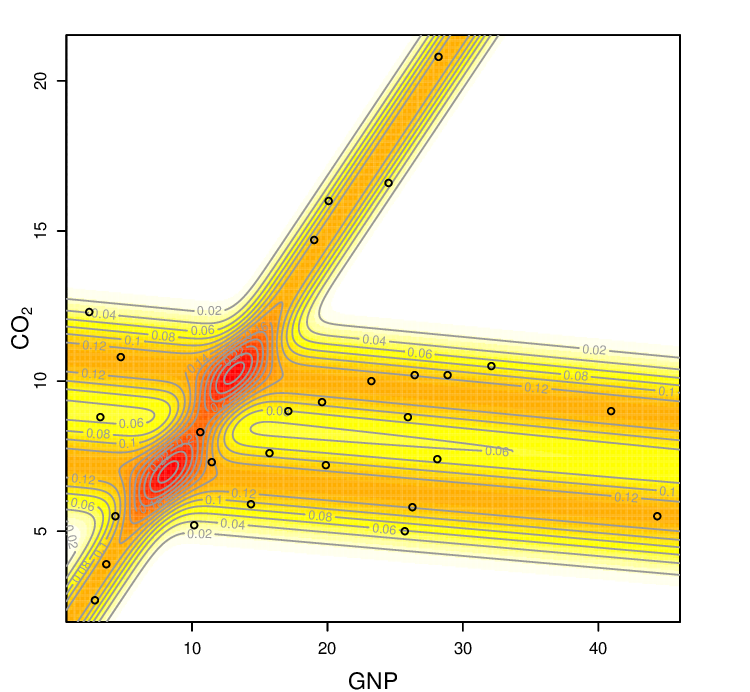}	
		\caption{Heat map of the conditional density of the outcome variable CO\textsubscript{2}, accounting for the gating and expert networks, the latter of which includes GNP as a covariate.}
		\label{Plot:CO2Density}
	\end{subfigure}%
	\caption{Scatterplots of GNP against CO\textsubscript{2} emissions for $n=28$ countries with three linear regression components from the optimal MoEClust model with equal variances and mixing proportions.\label{Plot:CO2Results}}
\end{figure}

Table \ref{Table:CO2Criterion} gives BIC and ICL values for the top model under each of the six special cases of the MoE framework. The chosen model had $G=3$, equal variances (i.e. the \textsf{E} constraint), equal mixing proportions, and GNP in the expert network; thus, this is an \emph{equal mixing proportion expert network MoE model}. This model maximised both the BIC and ICL criteria, and was also identified by the forward stepwise search described in Algorithm \ref{Alg:Stepwise}, starting from a $G=1$ model (BIC=$-163.90$), adding a component (BIC=$-163.16$), adding GNP to the expert network and changing to the \textsf{V} model type (BIC=$-157.20$), and finally adding a further component, constraining the mixing proportions, and changing back to the \textsf{E} model type (BIC=$-155.20$). Thereafter, neither adding a component nor adding GNP to the gating network improved the BIC. Code to reproduce both the exhaustive and stepwise searches using the \texttt{MoEClust} \textsf{R} package is given in Appendix \ref{Appendix:CO2code}.
\begin{table}[H]
	\caption{The MoEClust BIC and ICL values of the top models under the six MoE special cases for the CO\textsubscript{2} data. Each row is optimal with respect to $G$ and GPCM type, given the included covariates.}
	\label{Table:CO2Criterion}
	\centering
	\scriptsize
	\vspace{-1ex}
	\begin{tabular}[pos=center]{ l c c | c c | c c }
		\hline
		\centering
		Special Case & Gating & Expert & $G$ & GPCM & BIC & ICL\\
		\specialrule{.1em}{.01em}{.01em}
				Mixture Model &  & & 2 & \textsf{E} & $-163.16$ & $-163.91$\\
		Expert Network MoE Model & & GNP & $2$ & \textsf{V} & \textbf{$-157.20$} & \textbf{$-160.04$}\\
		Gating Network MoE Model & GNP & & $2$ & \textsf{E} & $-166.05$ & $-166.68$\\
		Full MoE Model & GNP & GNP & $2$ & \textsf{V} & $-159.25$ & $-161.47$\\
		Equal Mixing Proportion Mixture Model & Equal & & $2$ & \textsf{V} & $-165.19$ & $-184.71$\\
		Equal Mixing Proportion Expert Network MoE Model & Equal & GNP & $3$ & \textsf{E} & $\mathbf{-155.20}$ & $\mathbf{-159.06}$\\
		\hline
	\end{tabular}
\end{table}%
Repeating both the exhaustive and stepwise searches with the addition of a noise component for all models also failed to yield any model with an improved BIC. The fourth row of Table \ref{Table:CO2Criterion} corresponds to a \emph{full MoE}, with GNP included in both parts of~the model; its sub-optimal BIC highlights the benefits of the model selection approach.~The parameters of the optimal model are given in Table \ref{Table:CO2Params}. Its fit is exhibited in Figure~\ref{Plot:CO2Results}, which shows that the relationship between CO\textsubscript{2} and GNP is clustered around three~diff\-erent linear regression lines; one cluster of $8$ countries with a large slope value~and two equally-sized clusters, each with different intercepts but similar near-zero slope values. Clustering uncertainties, given by $\widehat{U}_i = \min_{g\,\in\,\left\{1,\ldots,\widehat{G}\right\}}\negmedspace\left(1 - \widehat{z}_{ig}\right)$, are also shown.
\begin{table}[H]
	\caption{Estimated parameters of the optimal MoEClust model fit to the CO\textsubscript{2} data.}
	\label{Table:CO2Params}
	\centering
	\scriptsize
	\vspace{-1ex}
	\begin{tabular}[pos=center]{ c | c c c}
		\hline
		\centering
		Parameter & Component 1 & Component 2 & Component 3\\
		\specialrule{.1em}{.01em}{.01em}
		Proportion & $\nicefrac{1}{3}$ & $\nicefrac{1}{3}$ & $\nicefrac{1}{3}$\\
		(Intercept) & $1.41$ & $7.29$ & $10.84$\\
		GNP & $0.68$ & $-0.04$ & $-0.04$\\
		$\sigma_g^2$ & $0.98$ & $0.98$ & $0.98$\\
		\hline
	\end{tabular}
\end{table}%
The optimal model contains GNP in the expert network and has constraints on~the component variances and mixing proportions. These are features of the MoEClust models which neither MoE nor GPCM models can fully accommodate. While \texttt{flexmix} and \texttt{mixtools} can fit the sub-optimal expert network MoE model in row four of Table \ref{Table:CO2Criterion}, with unequal variances and mixing proportions (which achieves the second highest BIC value), our initialisation strategy ultimately leads to the same or higher BIC estimates. Across $50$ random starts, BIC values of $-157.29$ and $-157.20$ are achieved using \texttt{flexmix} and \texttt{mixtools}, respectively. Among these random starts, BIC values as low as $-163.67$ are obtained. However, the \texttt{MoEClust} \textsf{R} package, with Algorithm \ref{Alg:InitEM} invoked, achieves a BIC of $-157.20$ using only a single initial partition. Using \texttt{MoEClust} without this initialisation strategy also yields the lower BIC value of $-163.67$. A further demonstration of the advantages of our initialisation strategy, using instead the optimal model for the the CO\textsubscript{2} data, is provided in Appendix~\ref{Appendix:CO2init}.

\subsection[Australian Institute of Sport (AIS) Data]{Australian Institute of Sport (AIS) Data}
\label{sec:AIS}
Various physical and hematological (blood) measurements were made on $102$ male and $100$ female athletes at the Australian Institute of Sport (AIS; \citealp{Cook1994}). The thirteen variables recorded in the study are detailed in Table \ref{Table:AISvariables}.
\begin{table}[H]
	\caption{Australian Institute of Sports data variables. The $p=5$ in the first column are hematological response variables and the others, the $r=8$ covariates, are physical measurements for the athlete.}
	\label{Table:AISvariables}
	\centering
	\scriptsize
	\vspace{-1ex}
	\begin{tabular}[pos=center]{l l l l}
		\hline
		Response & Description & Covariate & Description (Units)\\
		\specialrule{.1em}{.01em}{.01em}
		RCC & red cell count & BMI & body mass index (kg/m$^2$)\\
		WCC & white cell count & SSF & sum of skin folds (mm)\\
		Hc & Hematocrit & Bfat & body fat percentage (\%)\\
		Hg & Hemoglobin & LBM & lean body mass (kg)\\
		Fe & plasma ferritin & Ht & height (cm)\\
		&concentration&Wt & weight (kg)\\
		&&sex & a factor with levels: female, male\\
		&&sport & a factor with levels: Basketball, Field, Gymnastics, Netball, \\ 
		&&&Rowing, Swimming, Tennis, Track 400m, Track Sprint, Water Polo\\		
		\hline
	\end{tabular}
\end{table}
MoEClust models are used to investigate the clustering structure in the athletes' hematological measurements and investigate how covariates may influence these measurements and the clusters. Cluster allocations are initialised using model-based agglomerative hierarchical clustering. Results of the forward stepwise model search desc\-ribed in Algorithm \ref{Alg:Stepwise}, with all covariates considered for inclusion, are given in Appendix \ref{Appendix:AISstep}. The optimal model (BIC=$-4010.14$) is a $2$-component \textsf{EVE} \emph{equal mixing proportion expert network MoE model}, which thus has clusters of equal size, volume, and orientation, and unequal shape. Notably, the only covariate (sex), only enters in one part of the model, the expert network. 

The sub-optimal BIC values for the best model with all covariates in both parts of the model ($G=2$, \textsf{VVE}, BIC=$-4563.12$), the best model with all covariates in the expert network only ($G=1$, \textsf{EEE}, BIC=$-4234.79$), regardless of $\boldsymbol{\tau}$ being constrained or not, and the best model with all covariates in the gating network only ($G=2$,~\textsf{VEE}, BIC=$-4092.57$), highlight the need for the model selection strategy employed. As the optimal model uses the \textsf{EVE} constraints, it has $19$ covariance parameters; an otherwise exactly equivalent \textsf{VVV} model, having $30$ such parameters, yields a lower BIC of $-4056.19$, thus showcasing the benefits of the parsimonious covariance constraints. The difference of $11$ covariance parameters between these models is exactly one more than the number of regression parameters introduced by the expert network covariate. 

Subsequently, and purely for the purposes of comparing certain special cases of interest, an exhaustive search over a range of MoEClust models is conducted, with $G \in \{1,\ldots,9\}$. This is rendered feasible by only considering the covariates BMI and sex; allowing either, neither, or both to enter either, neither, or both of the gating and expert networks. Note that BMI is itself computed using the covariates measuring weight (Wt) and height (Ht). With $3$ permissible covariance parameterisations for the single component models (i.e. those without gating network covariates) and $14$ otherwise, $16$ possible combinations of gating and/or expert network covariate settings, and consideration also being given to $G>1$ models with equal mixing proportions, this still requires fitting $2,252$ MoEClust models. However, some $26$ spurious solutions were found (and ultimately discarded), particularly for higher values of $G$, in the sense that models with empty components or degenerate components with few observations reduced to equivalent models with fewer non-empty components (see Section \ref{sec:noise}). Table \ref{Table:AISCriterion} gives the BIC and ICL values of a selection of these fitted models, representing the optimal models for certain special cases of interest.
\begin{table}[H]
	\caption{The BIC and ICL values for a selection of MoEClust models fitted to the Australian Institute of Sports data. Rows 1 and 2 give the optimal models under settings available in \texttt{flexmix}; models without expert network covariates, using either the \textsf{VVV} or \textsf{VVI} covariance constraints. Among the more general MoEClust family, the last row gives the top model according to the ICL criterion and the remaining rows give the top models according to the BIC criterion for each of the six special cases of the MoE framework. Thus, rows 3 and 7 roughly correspond to the optimal models according~to \texttt{mclust}, with unequal and equal mixing proportions, respectively. For each model, its number of estimated parameters and its rank (according to BIC) among the full set of fitted models are~\mbox{also~given}.}
	\label{Table:AISCriterion}
	\centering
	\scriptsize
	\vspace{-1ex}
	\begin{tabular}[pos=center]{ c | c c | c c | c c | c}
		\hline
		\centering
		Rank (BIC) & Gating & Expert & $G$ & GPCM & BIC & ICL & No. Parameters\\
		\specialrule{.1em}{.01em}{.01em}
		$198$ & sex & & $2$ & \textsf{VVV} & $-4113.31$ & $-4121.32$ & $42$\\
		$880$ & sex & & $5$ & \textsf{VVI} & $-4319.85$ & $-4345.55$ & $58$\\
		\hline
		$293$ & && $2$ & \textsf{EVE} & $-4146.16$ & $-4201.61$ & $30$\\
		$3$ & & sex& $2$ & \textsf{EVE} & $-4015.35$ & $-4059.54$ & $40$\\
		$24$ & sex & & $3$ & \textsf{EVE} & $-4037.32$ & $-4066.66$ & $42$\\
		$2$ & BMI & sex& $2$ & \textsf{EVE} & $-4013.40$ & $-4074.11$ & $41$\\
		$269$ & Equal & & $2$ & \textsf{EVE} & $-4140.98$ & $-4192.21$ & $29$\\
		$1$ & Equal & sex & $2$ & \textsf{EVE} & $\mathbf{-4010.14}$ & $-4057.87$ & $39$\\
		\hline
		$26$ &BMI, sex & & $3$ & \textsf{EEE} & $-4038.64$ & $\mathbf{-4043.02}$ & $36$\\
		\hline
	\end{tabular}
\end{table}
Clearly, the inclusion of covariates improves the fit compared to GPCM models. Similarly, using GPCM covariance constraints improves the fit compared to standard Gaussian MoE models. In particular, it is notable that the optimal models using the \textsf{VVV} and \textsf{VVI} constraints only have covariates enter the gating network. This suggests that the parsimony afforded by the remaining GPCM settings somewhat offsets the number of regression parameters introduced to the expert network.

The top three models according to BIC all have $2$ components, the~\textsf{EVE} covariance constraints, and the covariate sex in the expert network; they differ only~in their treatment of the gating network. Models with~equal and unequal mixing propor\-tions, and with BMI as a gating concomitant, have zero,~one, and two associated gating network parameters, respectively. The optimal model has equal mixing proportions and was also identified above via Algorithm \ref{Alg:Stepwise}. The full MoE model with~BMI in the gating network and sex in the expert network is an interesting case as it does not fit the framework of \citet{Dang2015}, which assumes that when covariates enter the model, they enter in both parts. The best such model has `sex' in both networks ($G=2$, \textsf{EVE}) and achieves a BIC of $-4020.22$ with a corresponding rank of $8$.

\begin{figure}[H]
	\centering
	\includegraphics[width=\textwidth, keepaspectratio]{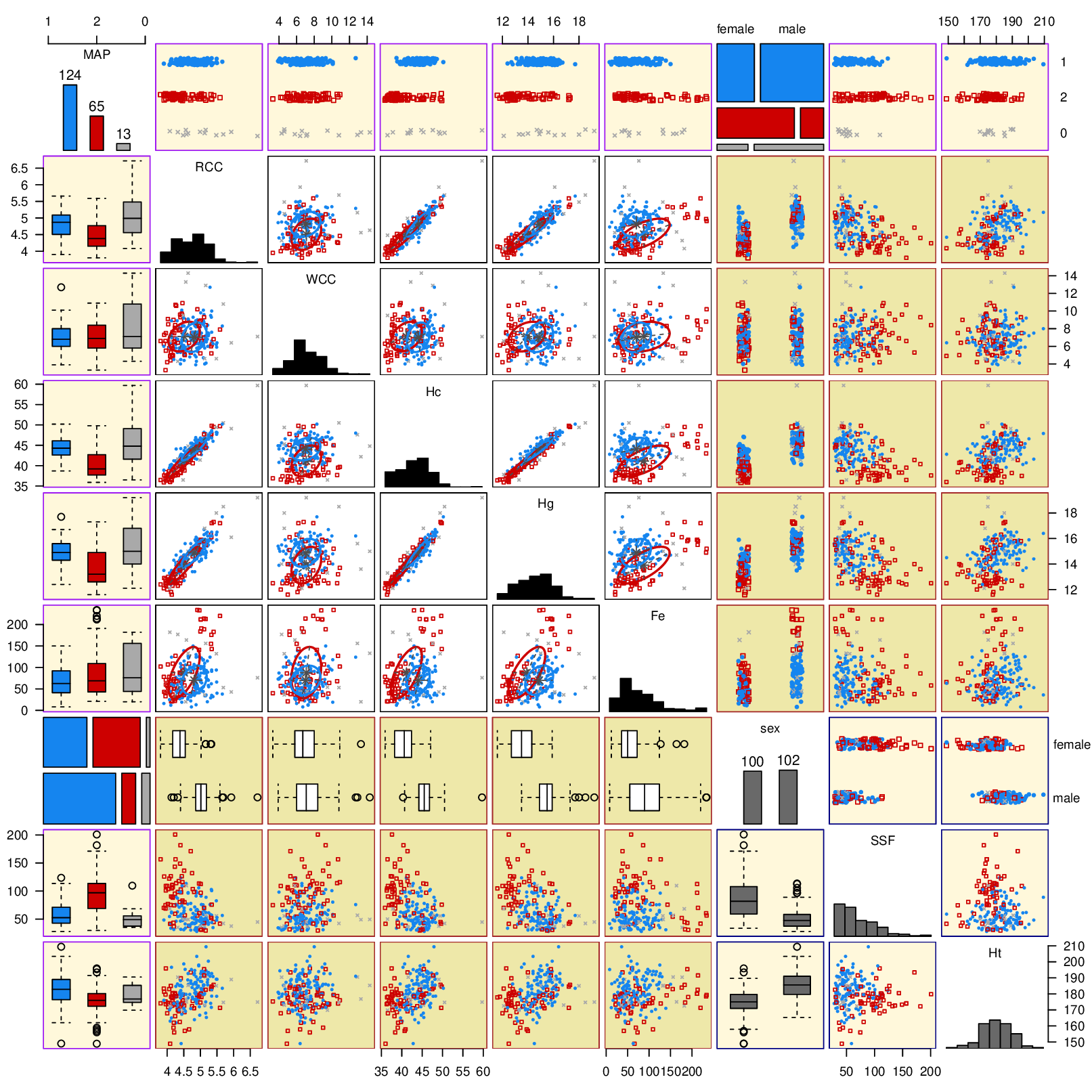}
	\caption{Generalised pairs plot for the optimal `full' MoEClust model fit to the AIS data, depicting pair\-wise relationships between the hematological response variables, the expert network covariate~sex, the gating concomitants SSF and Ht, and the MAP classification. Colours and plotting symbols correspond to the MAP classification: blue circles and red squares for the two Gaussian clusters; grey crosses for the $4$ female and $9$ male outlying observations assigned to the uniform noise component. Mosaic plots are used to depict two categorical variables, scatterplots are used for panels involving two continuous variables, and a mix of boxplots and jittered strip plots are used for mixed pairs.}
	\label{Plot:AISgpairs}
\end{figure}

Up to now, models with a noise component have not yet been considered for the~AIS data. Thus, another stepwise search is conducted, including a noise component for all candidate models and starting from a $G=0$ noise-only model  (see Appendix \ref{Appendix:AISstep}). Consideration was also given to both the GN and NGN settings, in \eqref{eq:noisegate} and \eqref{eq:noisenogate} respectively, where models included gating concomitants, and to models with equal/unequal mixing proportions for the non-noise components for models without gating concomitants. The optimal full MoE model thus found has two \textsf{EEE} Gaussian clusters and an additional noise component. The covariate `sex' enters the expert network (see Table \ref{Table:AISCoeff}). Both `SSF' and `Ht' enter the gating network, though not for the noise component, which has a constant mixing proportion ($\smash{\widehat{\tau}_0\approx0.08}$), as per the NGN setting in \eqref{eq:noisenogate}. Thus, the Gaussian clusters have equal volume, shape, and orientation, but unequal size. This model achieves a BIC value of $-3989.83$, which compares favourably to  the previously optimal model from Table \ref{Table:AISCriterion}, adding a noise component to a model otherwise identical to the optimal model from Table \ref{Table:AISCriterion} (BIC=$-3992.81$), and to models with a noise component but no stepwise selection of covariates (or no covariates at all).
\begin{table}[H]
	\caption{Coefficients of the expert network linear regressions for the $G=2$ Gaussian clusters in the optimal `full' MoEClust model (with an extra noise component and gating concomitants entering the~non-\-noise clusters only) fit to the AIS data, with female as the reference level for the explanatory variable~`sex'.}
	\centering
	\scriptsize
	\vspace{-1ex}
	\label{Table:AISCoeff}
	\begin{tabular}[pos=center]{c | c c c c c}
		\hline
		\centering
		\multirow{2}{*}{} & RCC & WCC & Hc & Hg & Fe\\
		\specialrule{.1em}{.01em}{.01em}
		\multicolumn{6}{c}{Cluster 1}\\
		\hline
		(Intercept) & $4.56$ & $6.89$ & $42.33$ & $14.08$ & $49.73$\\
		sexmale & $0.42$ & $0.12$ & $2.95$ & $1.30$ & $28.19$\\
		\hline
		\multicolumn{6}{c}{Cluster 2}\\
		\hline
		(Intercept) & $4.26$ & $6.93$ & $38.91$ & $13.11$ & $59.70$\\
		sexmale & $0.86$ & $0.59$ & $7.36$ & $2.80$ & $132.66$\\
		\hline
	\end{tabular}
\end{table}
\vspace{-1.5ex}
The gating network has an intercept of $10.58$ and slope coefficients of $0.04$ (SSF)~and $-0.08$ (Ht) with corresponding odds ratios of $1.04$ and $0.93$. Thus, higher SSF values increase the probability of belonging to the second Gaussian cluster, to which taller athletes are less likely to belong, and the probability~of belonging to the noise component is constant. Though every observation has its own mean parameter in the presence of expert covariates, given by the fitted values of the expert network (shown in Table \ref{Table:AISCoeff}), the means are summarised in Table \ref{Table:AISParams} by~the posterior mean of the fitted values of the model according to \eqref{eq:posteriormean}. The noise component is accounted for by $\overline{\bm{\mathcal{V}}}$, the $p$-dimensional centroid of the region used to estimate $V$:
\begin{equation}
\widehat{\boldsymbol{\mu}}_g=\frac{\sum_{i=1}^n\widehat{z}_{ig}\widehat{\mathbf{y}}_i}{\sum_{i=1}^n\widehat{z}_{ig}}=\frac{\sum_{i=1}^n\widehat{z}_{ig}\big(\sum_{g=1}^G\widehat{z}_{ig}\big(\widetilde{\mathbf{x}}_i^{\left(E\right)}\widehat{\boldsymbol{\gamma}}_g\big) + \widehat{z}_{i0}\overline{\bm{\mathcal{V}}}\big)}{\sum_{i=1}^n\widehat{z}_{ig}}.\label{eq:posteriormean}
\end{equation}
Given that there exists a binary variable, `sex', in the expert network for the optimal MoEClust model, there are effectively four Gaussian components plus an additional noise component. By virtue of the \textsf{EEE} constraint on the Gaussian components, all four components and thus both clusters in fact share the same covariance matrix. Components 1 and 2, corresponding to females and males in Cluster 1, share the same covariance matrix but differ according to their means. The same is true for females and males (Components 3 and 4) in Cluster 2. Table \ref{Table:AISParams} gives the means and average gates in terms of both components and clusters, as well as the common $\smash{\widehat{\boldsymbol{\Sigma}}}$ matrix.
\begin{table}[H]
	\caption{Estimated parameters of the $G=2$ Gaussian clusters in the optimal `full' MoEClust model fit to the AIS data (with an extra noise component and gating concomitants entering the non-noise clusters), with further splitting due to the binary covariate sex in the expert network, giving average gates and component means (for females and males) and the common \textsf{EEE} covariance matrix. While every observation has its own mean parameter, given by the fitted values of the expert network in Table \ref{Table:AISCoeff}, the means are summarised by the posterior mean of the model's fitted values, given by \eqref{eq:posteriormean}.}
	\label{Table:AISParams}
	\centering
	\scriptsize
	\vspace{-1ex}
	\begin{tabular}[pos=center]{c | c | c c | c | c c | c c c c c}
		\hline
		\centering
		& \multicolumn{3}{c|}{Cluster 1} & \multicolumn{3}{c|}{Cluster 2} & \multicolumn{5}{c}{\multirow{2}{*}{$\widehat{\boldsymbol{\Sigma}}$~(\textsf{EEE})}}\\
		\cline{2-7}
		& All & Female & Male & All & Female & Male &\multicolumn{5}{c}{}\\
		\specialrule{.1em}{.01em}{.01em}
		$\overline{\widehat{\tau}_g\negthinspace\left(\mathbf{x}_i\right)}$& $0.60$& $0.21$ & $0.39$ &$0.33$&$0.25$&$0.08$& RCC&WCC&Hc&Hg&Fe\\
		\cline{1-12}
		RCC  & $4.81$  &     $4.51$ &     $4.98$ &  $4.51$ &     $4.33$ &     $5.12$ &  $0.08$ & $0.08$ & $0.46$ & $0.15$ & $-0.83$\\
		WCC  & $7.02$  &     $6.95$ &     $7.06$ &  $7.10$ &     $6.96$ &     $7.57$ &         & $2.50$ & $0.60$ & $0.21$ & $5.12$\\
		Hc   & $44.06$ &    $41.79$ &    $45.35$ & $41.14$ &    $39.61$ &    $46.29$ &         &        & $3.84$ & $1.33$ & $-7.55$\\
		Hg   & $14.88$ &    $13.94$ &    $15.51$ & $13.91$ &    $13.32$ &    $15.90$ &         &        &        & $0.57$ & $-1.05$\\
		Fe   & $70.18$ &    $53.05$ &    $79.87$ & $87.84$ &    $58.96$ &   $184.67$ &         &	    &        &        & $821.68$\\
		\hline
	\end{tabular}
\end{table}
\vspace{-1.5ex}
Though the plots in Figure \ref{Plot:CO2Results} are suitable for univariate data with a single continuous expert network covariate, visualising MoEClust results for multivariate data~with $r>1$ mixed-type covariates constitutes a much greater challenge. For the optimal full MoE model fit to the AIS data, the data and clustering results are shown using a generalised pairs plot \citep{Emerson2013} in Figure \ref{Plot:AISgpairs}. This plot depicts the pairwise relationships between the hematological response variables, the included gating and expert network covariates, and the MAP classification, coloured according to the MAP classification. The marginal distributions of each variable are given along the diagonal. For the hematological response variables, ellipses with axes related to the within-cluster covariances are drawn. For the purposes of visualising Figure \ref{Plot:AISgpairs}, owing to the presence~of an expert network covariate in the fitted model, the multivariate Gaussian ellipses in panels depicting two response variables are centred on the posterior mean of the fitted values, as described~in \eqref{eq:posteriormean}. Their volume, shape, and orientation are also modified for the same reason:~they are derived by adding the extra variability in the fitted values (weighted by $\smash{\widehat{\mathbf{Z}}}$) to $\smash{\widehat{\boldsymbol{\Sigma}}_g}$. Thus, the depicted ellipses do not conform to the \textsf{EEE} covariance constraints of the optimal~model.\enlargethispage{\baselineskip}\vspace{-1ex}
It is clear from Figure \ref{Plot:AISgpairs} that the variables `Hematocrit' (Hc), `Hemoglobin' (Hg), and `plasma ferritin concentration' (Fe), and the gating network concomitants `SSF' and `Ht', are driving much of the separation between the clusters. On the other hand, the expert network covariate `sex' is driving separation within the Gaussian clusters. The correspondence between the MAP classification and the sex label is notably poor, with an adjusted Rand index \citep{Hubert1985} of just $0.11$. This index is higher for models where sex does not enter the expert network, especially when it instead enters the gating network, though such fitted models all have sub-optimal BIC values (see Table \ref{Table:AISCriterion}). This is because, under the optimal model, the athletes' size in terms of their SSF and height measurements, rather than their sex, influences the probability of cluster membership, and athletes are divided by sex within each cluster rather than the clusters necessarily capturing their~sex.

Indeed, Table \ref{Table:AISCoeff} implies that males, on average, have elevated levels of all five blood measurements~in both Gaussian clusters. However, the magnitude of this effect is more pronounced in Cluster 2, related to athletes with higher average~SSF measurements (a proxy for body fat) and lower average height. Interestingly, Figure \ref{Plot:AISgpairs} also shows that females have higher average SSF measurements and lower average height; this may explain why there are more males than females in Cluster 1, and the reverse in Cluster 2, given the signs of the gating network coefficients for SSF ($0.04$) and Ht ($-0.08$).\vspace{-0.25em}

\section[Discussion]{Discussion}
\label{sec:Discussion}
The development of a suite of MoEClust models has been outlined, clearly demonstrating the utility of mixture of experts models for parsimonious model-based clustering where covariates are available. A novel means of visualising such models has also been proposed. The ability of MoEClust models to jointly model the response variable(s) and related covariates provides deeper and more principled insight into the relations between such data in a mixture-model based analysis, and provides a principled method for both creating and explaining the clustering, with reference to information contained in covariates. Their demonstrated use to cluster observations and appropriately capture heterogeneity in cross-sectional data provides only a glimpse of their potential flexibility and utility in a wide range of settings. Indeed, given that general MoE models have been used, under different names, in several fields, including but not limited to statistics \citep{Grun2007,Grun2008}, biology \citep{Wang1996}, econometrics \citep{Wang1998}, marketing \citep{Wedel2012}, and medicine \citep{Thompson1998}, MoEClust models could prove useful in many domains. 

Improvement over GPCM models has been introduced by accounting for external information available in the presence of potentially mixed-type covariates. Similarly, improvement over Gaussian mixture of experts models which incorporate fixed covariates has been introduced by allowing GPCM family covariance structures in the component densities. MoEClust models are thus Gaussian parsimonious MoE models where the size, volume, shape, and/or orientation can be equal~or unequal across components. MoEClust models have been further extended to accommodate the presence of an additional uniform noise component to capture departures from Gaussianity, in such a way that observations are smoothly classified as belonging to Gaussian clusters or as outliers. In particular, two means of doing so have been proposed for models which include gating concomitants. Due to sensitivity of the final solution obtained by the EM algorithm to the initial values, an iterative reallocation procedure based~on the Mahalanobis distance has been proposed~to mitigate against convergence to suboptimal local maxima for models with expert network covariates.~This initialisation algorithm converges quickly and also speeds up convergence of the EM algorithm itself.

Previous parsimonious Gaussian mixtures of experts \citep{Dang2015} accommodated only the cases where all covariates enter the expert network MoE model, or the full MoE model with the restriction that all covariates enter both parts of the model. MoEClust constitutes a unifying framework whereby different subsets of covariates can enter either, neither, or both the gating and/or expert networks of Gaussian parsimonious MoE models. Considering the standard mixture model (with no dependence on covariates), or the expert network MoE model, with the equal mixing proportion constraint expands the model family further.

On a cautionary note, care must be exercised in choosing how and where covariates enter when a MoEClust model is used as a clustering tool, as the interpretation of the analysis fundamentally depends on where covariates enter, which of the six special cases of the MoE framework is invoked, and on which GPCM constraints are employed. To this end, a novel greedy forward stepwise search algorithm has been employed for model/variable selection purposes. This strategy has the added advantage of introducing additional parsimony, by potentially reducing the number of regression parameters in the gating and/or expert networks.

Gating network MoEClust models may be of particular interest to users of GPCMs; while concomitants influence the probability of cluster membership, the correspondence thereafter between component densities and clusters has the same interpretation as in standard GPCMs. When covariates enter the component densities, we warn that observations with very different response values can be clustered together, because they are being modelled using the same GLM; similarly, regression distributions with distinct parameters do not necessarily lead to well-separated~clusters. 

MoEClust models allow the number of parameters introduced by gating and expert network covariates to be offset by a reduction in the number of covariance parameters. This is particularly advantageous when model selection is conducted using the BIC or ICL, which include a penalty term based on the parameter count. Thus, MoEClust models may tend to favour including covariates more than standard Gaussian MoE models would. This is particularly true for explanatory variables in the expert network, which tend to necessitate more regression parameters ($Gp$) than concomitant variables in the gating network ($G-1$) per additional continuous covariate or level of categorical covariates included. Thus, in cases where a MoE model might elect to include a concomitant variable in the gating network, a MoEClust model with fewer covariance parameters may elect to include it as an explanatory expert network variable instead. While this does lead to a better fit, it can complicate interpretation. 

Possible directions for future work in this area include investigating the utility of nonparametric estimation of the gating network  \citep{Young2010}, as well as exploring the use of regularisation penalties in the gating and expert networks to help with variable selection when the number of covariates $r$ is large. By shrinking some coefficient estimates to zero, this sparsity could help with relaxing the assumptions that covariates affect all components and, in the case of the expert network, the assumption that covariates affect the means of all $p$ dependent variables. Regularisation in another, Bayesian sense, by specifying a prior on the component variances/covariances in the spirit of \citet{Fraley2007}, and/or component regression parameters, could also prove useful for avoiding spurious solutions due to computational singularities, as described in Section \ref{sec:noise}. MoEClust models could also be developed in the context of hierarchical mixtures of experts \citep{Jordan1994}, and/or extended to the supervised or semi-supervised model-based classification settings, where some or all observations are labelled.

Beyond the family of GPCM constraints, MoEClust models could be extended to avail of parsimonious factor-analytic covariance structures for high-dimensional data \citep{McNicholas2008,McNicholas2010b}. These could be incorporated into Gaussian mixture of experts models using residuals in an equivalent fashion to Section \ref{sec:MoEClustInference} above. Similarly, MoEClust models could benefit from the heavier tails of the multivariate $t$-distribution, and the robustness to outliers it affords, by considering the associated $t$EIGEN family of covariance constraints \citep{Andrews2012}. However, the inclusion of a uniform noise component has the advantage of drawing a clearer distinction between observations belonging to clusters or designated as outliers.

\section*{Acknowledgements}
This work was supported by the Science Foundation Ireland funded Insight Centre for Data Analytics in University College Dublin under grant number SFI/12/RC/2289\_P2.

\setlength{\bibsep}{5.875pt}
\bibliographystyle{myplainnat}\interlinepenalty=10000
\bibliography{MoEClustBib}

\let\appendixpagenameorig\appendixpagename
\renewcommand{\appendixpagename}{\Large\appendixpagenameorig}
\makeatletter
\renewcommand\section{\@startsection{section}{1}{\z@}%
	{-3.5ex \@plus -1ex \@minus -.2ex}%
	{2.3ex \@plus.2ex}%
	{\normalfont\large\bfseries}}
\makeatother
\clearpage

\begin{appendices}
	\renewcommand\thetable{\thesection.\arabic{table}}
	\renewcommand\thefigure{\thesection.\arabic{figure}}
	\vspace{-1ex}
	\section[CO\textsubscript{2} Code Examples]{CO\textsubscript{2} Data: Code Examples}
	\label{Appendix:CO2code}
	\setcounter{table}{0}
	\setcounter{figure}{0}
	
	\definecolor{codegreen}{rgb}{0,0.6,0}
	\definecolor{codegray}{rgb}{0.5,0.5,0.5}
	\definecolor{codepurple}{rgb}{0.58,0,0.82}
	\definecolor{backcolour}{rgb}{0.95,0.95,0.92}
	\lstdefinestyle{mystyle}{
		backgroundcolor=\color{backcolour},   
		commentstyle=\color{codegreen},
		keywordstyle=\color{blue},
		numberstyle=\tiny\color{codegray},
		stringstyle=\color{codepurple},
		basicstyle=\linespread{1.2}\ttfamily\small,
		literate=
			{~}{{\fontfamily{ptm}\selectfont \textasciitilde}}1
			{<-}{{\ttfamily <-}}1,
		breakatwhitespace=false,         
		breaklines=true,                 
		captionpos=t,                    
		keepspaces=true,                 
		numbers=left,                    
		numbersep=5pt,                  
		showspaces=false,                
		showstringspaces=false,
		showtabs=false,     
		tabsize=2,
		mathescape=true,
		otherkeywords = {!,!=,*,\&,\%/\%,\%*\%,\%\%},
		morecomment=[l]{\#}
	}
	
	\lstset{style=mystyle, keywords={MoE_clust, MoE_compare, MoE_stepwise, library, data, list, TRUE}}
	
	Code to reproduce both the exhaustive (Listing \ref{list:exhaustive}) and greedy forward stepwise (Listing \ref{list:stepwise}) searches for the CO\textsubscript{2} data described in Section \ref{sec:CO2}, using the \texttt{MoEClust} \textsf{R} package \citep{MoEClustR2021}, is provided below. The code in Listing \ref{list:exhaustive} can be used to reproduce the results in Table \ref{Table:CO2Criterion}.
\begin{lstlisting}[numbers=none, caption={Exhaustive search \textsf{R} code for the CO\textsubscript{2} data.}, label={list:exhaustive}]
	library(MoEClust)
	data(CO2data)
	CO2   <- CO2data$\$$CO2
	GNP   <- CO2data$\$$GNP

 # Fit models under the 6 special cases of the MoE framework
	m1    <- MoE_clust(CO2, G=1:9)
	m2    <- MoE_clust(CO2, G=2:9, gating= ~ GNP)
	m3    <- MoE_clust(CO2, G=1:9, expert= ~ GNP)
	m4    <- MoE_clust(CO2, G=2:9, gating= ~ GNP, expert= ~ GNP)
	m5    <- MoE_clust(CO2, G=2:9, equalPro=TRUE)
	m6    <- MoE_clust(CO2, G=2:9, equalPro=TRUE, expert= ~ GNP)
 
 # Collate results and rank (by BIC) only the 6 optimal models
	res   <- list(m1=m1, m2=m2, m3=m3, m4=m4, m5=m5, m6=m6)
	(comp <- MoE_compare(res, optimal.only=TRUE))\end{lstlisting}
\begin{lstlisting}[numbers=none, caption={Stepwise search \textsf{R} code for the CO\textsubscript{2} data.}, label={list:stepwise}]
	library(MoEClust)
	data(CO2data)
	CO2   <- CO2data$\$$CO2
	GNP   <- CO2data$\$$GNP

 # Conduct a stepwise search
  (mod1 <- MoE_stepwise(CO2, GNP))
  
 # Conduct a stepwise search for models with a noise component
  (mod2 <- MoE_stepwise(CO2, GNP, noise=TRUE))
  
 # Compare both sets of results to choose the optimal model
  (best <- MoE_compare(mod1, mod2, optimal.only=TRUE)$\$$optimal)
\end{lstlisting}
	
	\section[CO\textsubscript{2} Initialisation]{CO\textsubscript{2} Data: EM Initialisation}
	\label{Appendix:CO2init}
	\setcounter{table}{0}
	\setcounter{figure}{0}
	
\begin{figure}[H]
		\centering
		\begin{subfigure}[t]{.49\linewidth}
			\centering
			\includegraphics[width=\textwidth, keepaspectratio]{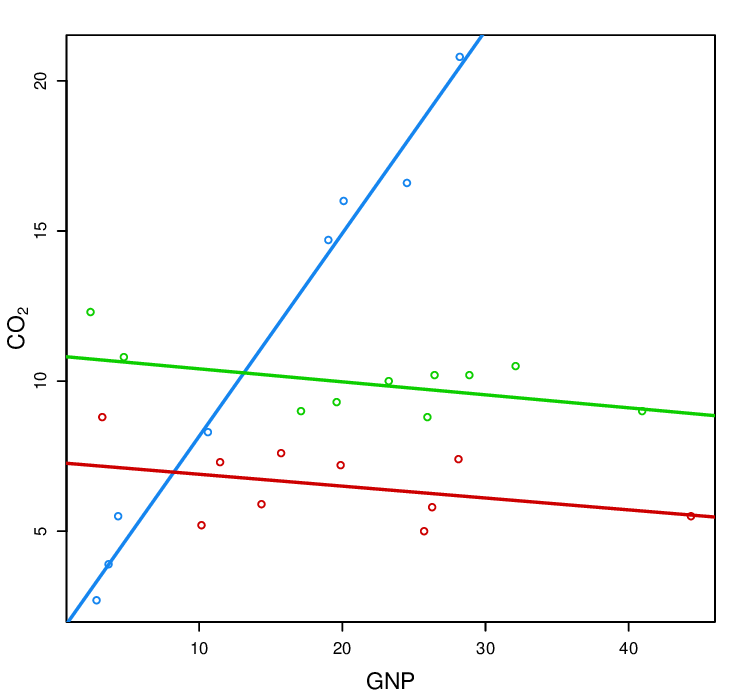}	
			\caption{With Algorithm \ref{Alg:InitEM} invoked for initialisation, achieving a BIC value of $-155.20$.}
			\label{Plot:CO2Init}
		\end{subfigure}\hfill%
		\begin{subfigure}[t]{.49\linewidth}
			\centering
			\includegraphics[width=\textwidth, keepaspectratio]{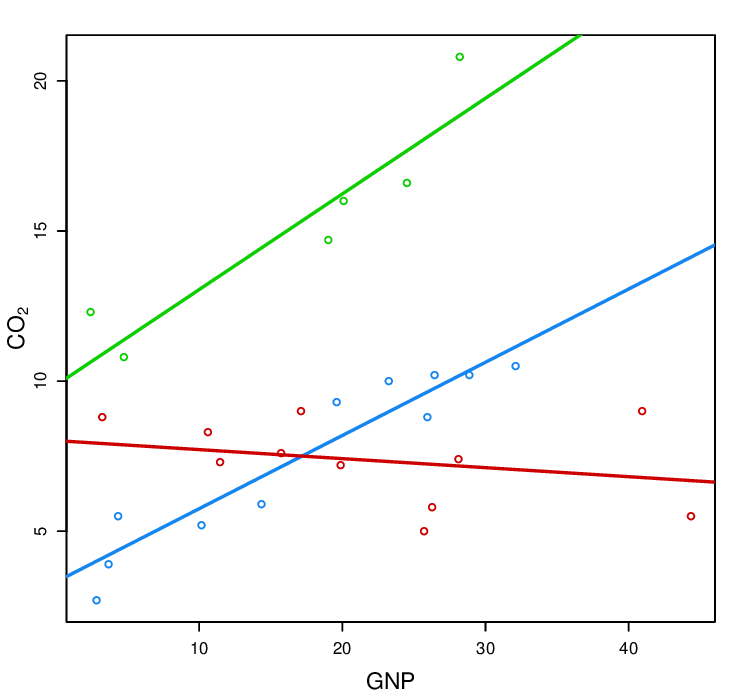}	
			\caption{Without Algorithm \ref{Alg:InitEM} invoked for initialisation, achieving a BIC value of $-161.06$.}
			\label{Plot:CO2NoInit}
		\end{subfigure}%
		\caption{Scatterplots of GNP against CO\textsubscript{2} emissions for $n=28$ countries showing $G=3$ coloured linear regression components from MoEClust models with equal variances and mixing proportions, with (\subref{Plot:CO2Init}) and without (\subref{Plot:CO2NoInit}) the initialisation strategy described in Algorithm \ref{Alg:InitEM} invoked.\label{Plot:CO2Initialisation}}
	\end{figure}

	The regression lines for the optimal $G=3$ equal mixing proportion expert network MoEClust model with equal component variances and the explanatory variable GNP fitted to the CO\textsubscript{2} data, with and without the initial partition obtained by model-based agglomerative hierarchical clustering being passed through Algorithm \ref{Alg:InitEM}, are shown in Figure \ref{Plot:CO2Initialisation}. A BIC value of $-155.20$ is achieved after $21$ EM iterations (starting, after $6$ iterations of our proposed initialisation strategy, from a log-likelihood of $-66.01$) compared to a value~of $-161.06$ after $28$ EM iterations without Algorithm \ref{Alg:InitEM} (starting from a log-likelihood of $-76.39$). While the models differ only in terms of the initialisation strategy employed, Table \ref{Table:CO2Criterion} shows that the model would not have been identified as optimal according to the BIC criterion had Algorithm \ref{Alg:InitEM} not been used. The superior solution in Figure \ref{Plot:CO2Init} has one cluster with a steep slope and two clusters with near-zero slopes but different intercepts. Notably, supplying $100$ random starts to Algorithm \ref{Alg:InitEM} did not yield an improved BIC in any instance. Similarly, the optimal BIC value obtained by supplying $100$ random starts to the EM algorithm  directly was greater than $-161.06$ but still less than $-155.20$.  	
			
	\section[AIS Stepwise Model Search]{AIS Data: Stepwise Model Search}
	\label{Appendix:AISstep}
	\setcounter{table}{0}
	\setcounter{figure}{0}
	
	For the AIS data, Table \ref{Table:AISstep} gives the results of the greedy forward stepwise model selection strategy described in Algorithm \ref{Alg:Stepwise}, showing the action yielding the best improvement in terms of BIC for each step. This forward search is less computationally onerous than its equivalent in the backwards direction. A $2$-component \textsf{EVE} \emph{equal mixing proportion expert network MoE} model is chosen, in which the mixing proportions are constrained to be equal and sex enters the expert network. This same model was identified after~an exhaustive search over a range of $G$ values, the full range of GPCM covariance constraints, and every possible combination of the BMI and sex covariates in the gating and expert networks (see Table \ref{Table:AISCriterion}). Note, however, that the remaining covariates in Table \ref{Table:AISvariables} are also considered for inclusion here. 
	
	To give consideration to outlying observations departing from the prevailing pattern of Gaussianity, a separate stepwise search is conducted, starting from a $G=0$ noise-only model, with all candidate models having an additional noise component. Thus, a distinction is made between the model found by following the steps shown in Table \ref{Table:AISstep} with $G=2$ \textsf{EVE} Gaussian components, and the model found by the second stepwise search described in Table \ref{Table:AISstepNoise} with three, of which two are \textsf{EEE} Gaussian and one~is uniform. Ultimately, the model with the noise component identified in Table \ref{Table:AISstepNoise} is chosen, based on its superior BIC. Aside from the noise component, it similarly includes `sex' in the expert network, but differs in its treatment of the gating network and the GPCM constraints employed for the Gaussian clusters. It is a \emph{full MoE} model where the Gaussian clusters have equal volume, shape, and orientation, the expert network includes the covariate `sex', and the both `SSF' and `Ht' influence the probability of belonging to the Gaussian clusters but not the additional noise component, as per \eqref{eq:noisenogate}.
	\begin{table}[H]
		\renewcommand\thesubtable{\alph{subtable}}
		\caption{Results of the forward stepwise model selection algorithm applied to the AIS data where candidate models do not include a noise component. All covariates in Table \ref{Table:AISvariables} are considered for inclusion in both parts of the model. The optimal action and associated BIC value is detailed for each step. The resulting models are described in terms of the number of Gaussian components $G$, the GPCM constraints used, and the treatment of the gating and expert networks.}
		\label{Table:AISstep}
		\centering
		\scriptsize
		\vspace{-1ex}
		\begin{tabular}[pos=center]{ c c | c c | c c | c}
			\hline
			\centering
			Step & Optimal Action & $G$ & GPCM & Gating & Expert & BIC\\
			\specialrule{.1em}{.01em}{.01em}
			1 & --- & $1$ & \textsf{EEE} & --- & & $-4202.79$\\
			2 & Add explanatory variable (Expert) & $1$ & \textsf{EEE} & --- & sex & $-4050.64$\\
			3 & Add component and constrain mixing proportions & $2$ &  \textsf{EVE} & Equal & sex & $-4010.14$\\
			4 & Stop & $2$ & \textsf{EVE} & Equal & sex & $-4010.14$\\
			\hline
		\end{tabular}
	\end{table}	
	\begin{table}[H]
		\renewcommand\thesubtable{\alph{subtable}}
		\caption{Results of the forward stepwise model selection algorithm applied to the AIS data where all candidate models explicitly include a noise component. All covariates in Table \ref{Table:AISvariables} are considered for inclusion in both parts of the model. The optimal action and associated BIC value is detailed for each step. The resulting models are described in terms of the number of Gaussian (i.e. non-noise) components $G$, the GPCM constraints used, and the treatment of the gating and expert networks. When gating concomitants are included, the chosen models here correspond to the NGN setting~in \eqref{eq:noisenogate}. Thus, the noise component's mixing weight is constant and independent of the included concomitants.}
		\label{Table:AISstepNoise}
		\centering
		\scriptsize
		\vspace{-1ex}
		\begin{tabular}[pos=center]{ c c | c c | c c | c}
			\hline
			\centering
			Step & Optimal Action & $G$ & GPCM & Gating & Expert & BIC\\
			\specialrule{.1em}{.01em}{.01em}
			1 & --- & $0$ & --- & --- & --- & $-4869.82$\\
			2 & Add component & $1$ & \textsf{EEE} & & 	 & $-4149.46$\\
			3 & Add explanatory variable (Expert) & $1$ & \textsf{EEE} &  & sex & $-4013.55$\\
			4 & Add component & $2$ & \textsf{EVE} & & sex & $-3992.81$\\
			5 & Add concomitant (Gating) & $2$ & \textsf{EVE} & NGN: SSF & sex & $-3990.09$\\
			6 & Add concomitant (Gating) & $2$ & \textsf{EEE} & NGN: SSF, Ht & sex & $-3989.83$\\
			7 & Stop & $2$ & \textsf{EEE} & NGN: SSF, Ht & sex & $-3989.83$\\
			\hline
		\end{tabular}
	\end{table}	
\end{appendices}

\end{document}